%%%%%%%%%%%%%%%%%%%%%%%  filename = stt.lanl.ellis.tex  %%%%%%%%%%%%%%%%%%%%%%%
%                                                                             %
%  This source file is written in TeX and uses AMSTeX macros.                 %
%                                                                             %
%%%%%%%%%%%%%%%%%%%%%%%%%%%%%%%%%%%%%%%%%%%%%%%%%%%%%%%%%%%%%%%%%%%%%%%%%%%%%%%

\nopagenumbers
\input amstex

\def\degree{\mathaccent"17 {}}

\def\Ghat{\hat G}
\def\epshat{\hat \epsilon}
\def\bdhat{\, @, @, @, @, @,{\hat {\! @! @! @! @! @!{\bold d}}}}
\def\omegahat{\hat \omega}
\def\tauhat{\hat \tau}

\def\Ehat{\hat E}
\def\Thetahat{\hat \Theta}
\def\Phihat{\hat \Phi}
\def\Psihat{\hat \Psi}
\def\ghat{\hat g}

\def\bdo{\hbox{$\, @, @, @, @, @,{\mathaccent"17
                 {\! @! @! @! @! @!{\bold d}}}$}}

\def\Ao{\mathaccent"17 A}
\def\Co{\mathaccent"17 C}
\def\Eo{\mathaccent"17 E}
\def\Fo{\mathaccent"17 F}
\def\go{\mathaccent"17 g}
\def\Go{\mathaccent"17 G}
\def\mo{\mathaccent"17 m}
\def\Po{\mathaccent"17 P}
\def\Ro{\mathaccent"17 R}

\def\tauo{\mathaccent"17 \tau}
\def\phio{\mathaccent"17 \phi}

\def\omegao{\mathaccent"17 \omega}

\def\rbraceo{\} \, \mathaccent"17 {}}
\def\Gammao{\mathaccent"17 \Gamma}
\def\Deltao{\mathaccent"17 \Delta}
\def\Thetao{\mathaccent"17 \Theta}
\def\Phio{\mathaccent"17 \Phi}
\def\Psio{\mathaccent"17 \Psi}

\def\pdot{\dot p}
\def\qdot{\dot q}
\def\zetadot{\dot \zeta}
\def\phidot{\dot \phi}

\def\pbdotdot{{\bold {\,\ddot {\! \text{\it p}}}}}
\def\Pbdot{{\bold {\,\,\dot {\!\! \text{\it P}}}}}
\def\bdrdotdotdot{\dot {\ddot {\text{\bf r}}}}

\def\dom{\operatorname{dom}}
\def\sgn{\operatorname{sgn}}

\def\lotimes{\! \otimes}
\def\lwedge{\! \wedge}

\def\dbar{{\bar d}}
\def\Kbar{{\bar K}}
\def\kappabar{{\bar \kappa}}
\def\Lbar{{\bar L}}
\def\lambdabar{{\bar \lambda}}
\def\Mbar{{\bar M}}
\def\mubar{{\bar \mu}}
\def\nubar{{\bar \nu}}
\def\tbar{{\bar t}}

\def\L{\bigl}     \def\R{\bigr}
\def\l{\left}     \def\r{\right}

\def\[[{[\![}     \def\]]{]\!]}

\documentstyle{amsppt}
\loadbold
\loadeusm
\TagsOnRight

\magnification=\magstephalf

\hoffset=.25truein
\hsize=6.0truein
\vsize=9.0truein
\voffset=-.25truein         % This is here because AMSTeX has started
                            % putting a 9 inch page 1/4 inch too low.
\parindent=20pt

\topmatter
\title Space-Time-\!-Time \endtitle
\author Homer G. Ellis\endauthor

\address
\vskip -30pt
\rm
$$
\alignat 2
\text{First version:  }& \;\;\;\;\;@,@,@,\text{August, 1989}& \\
\vspace{-3pt}
\text{Revised:  }& \text{September, 2001}& \\
\vspace{-3pt}
\text{Revised:  }& \;\;\;\;\;\;\;\;\,@!@!\text{April, 2002}&
\endalignat
$$
\vskip -5pt
Homer G. Ellis\newline
\indent Department of Mathematics\newline
\indent University of Colorado at Boulder\newline
\indent 395 UCB\newline
\indent Boulder, Colorado  80309-0395\newline\newline
\indent Telephone: (303) 492-7754 (office); (303) 499-4027 (home)\newline
\indent Email:  ellis\@euclid.colorado.edu\newline
\indent Fax: (303) 492-7707
\endaddress 

\abstract{\it
Space-time-\!-time is a natural hybrid of Kaluza's five-dimensional geometry
and Weyl's conformal space-time geometry.  Translations along the secondary
time dimension produce the electromagnetic gauge transformations of
Kaluza--Klein theory and the metric gauge transformations of Weyl theory,
quantitatively related as Weyl postulated.  Geometrically, this phenomenon
resides in an exponential-expansion producing ``conformality constraint'',
which replaces Kaluza's ``cylinder condition'' and is applicable to metrics of
all dimensionalities and signatures.  The de Sitter space-time metric is
prototypically conformally constrained; its hyper-de Sitter analogs of
signatures $+++-+$ and $+++--$ describe \text{space-time-\!-time} vacua.  The
curvature tensors exhibit in \text{space-time-\!-time} a wealth of
``interactions'' among geometrical entities with physical interpretations.
Unique to the conformally constrained geometry is a sectionally isotropic,
ultralocally determined ``residual curvature'', useful in construction of an
action density for field equations.  A \text{space-time-\!-time} geodesic
describes a test particle whose rest mass $\mo$ and electric charge $q$ evolve
according to definite laws.  Its motion is governed by four apparent forces:
the Einstein gravitational force proportional to $\mo$, the Lorentz
electromagnetic force proportional to $q$, a force proportional to $\mo$ and to
the electromagnetic four-potential, and a force proportional to $q^2 / \mo$ and
to the gradient of $\ln \phi$, where the scalar field $\phi$ is essentially the
\text{space-time-\!-time} residual radius of curvature.  The particle appears
suddenly at an event ${\eusm E}_1$ with $q = -\phi ({\eusm E}_1)$ and vanishes
suddenly at an event ${\eusm E}_2$ with $q = \phi ({\eusm E}_2)$.  At
${\eusm E}_1$ and ${\eusm E}_2$ the $\phi$-force infinitely dominates the
others, causing ${\eusm E}_1$ and ${\eusm E}_2$ to occur near where $\phi$ has
an extreme value; application to the modeling of orbital transitions of atomic
electrons suggests itself.  The equivalence of a test particle's inertial mass
and its passive gravitational mass is a consequence of the gravitational
force's proportionality to $\mo$.  No connection is apparent between $\mo$ and
active gravitational mass or between $q$ and active electric charge, nor does
the theory seem to require any.  Justification for applying the name
\text{``space-time-\!-time''} whether the signature be $+++-+$ or $+++--$ lies
in a construction which, applied to Euclidean spheres, produces the de Sitter
manifold and its time coordinate $t$ (``space's time''), and, when applied to
Minkowskian spheres, produces the hyper-de Sitter manifolds and their new
coordinate $\zeta$ (``space-time's time'').  If \text{space-time-\!-time}
metrics of the two signatures are placed on equal footing by complexification
of $\zeta$, the expanded geometry presents new elements which beg to be linked
to quantum mechanical phase phenomena.  The forging of such a link will allow
one ultimately to say, not that geometry has been quantized, but that the
quantum has been geometrized.
}\endabstract

\endtopmatter

\document

\vskip -10pt

\noindent {\bf I.  Introduction}
\vskip 10pt

Impelled by convictions about the nature of time \cite{1}, I have pursued the
prospect that manifolds bearing ``conformally constrained'' metrics can serve
as realistic models of physical systems in which gravity, electromagnetism, and
other phenomena manifest themselves.  This paper presents some of the products
of that ongoing pursuit.

Roughly, a conformally constrained metric is one for which there is a vector
field $\xi$ such that the lengths of vectors Lie transported by $\xi$ are
conformally expanded if those vectors are orthogonal to $\xi$, but are left
unchanged if those vectors are parallel to $\xi$ \cite{2}; the de Sitter metric
is the prototype.  The geometry of five-dimensional manifolds carrying such
metrics is a natural hybrid of the five-dimensional Kaluza geometry, with its
distinguished Killing vector field that ``isometrically constrains'' the metric
\cite{3}, and the four-dimensional Weyl geometry, with its multiplicity of
conformally related metrics and the associated gauge forms \cite{4}.  This
Kaluza--Weyl offspring is an evolutionary improvement in that it retains and
enhances the most useful characters of its progenitors while attenuating to
benign and useful form those that have caused difficulty.  Most notably, it
retains both the Kaluza unification of gravity with electromagnetism and the
Weyl association of metrical gauge changes (multiplications of the metric by
conformal factors) with electromagnetic gauge changes (additions of gradients
to the electromagnetic potential).  Also, it converts the objectionable
nonintegrability of length transference in the Weyl geometry to integrability
without sacrificing the principle that length, because it is a comparative
measure, depends on designation of a standard at each point, that is, on choice
of a gauge.  In the process it lends to the fifth dimension an essential
significance that the Kaluza geometry fails to provide.

The picture that emerges from application of this hybrid geometry to the
modeling of physical systems has in it some rather unexpected representations
of elementary physical phenomena, quantum phenomena included.  Because the 
models are clearly defined, with little room for ambiguity in their
interpretations, these representations appear to be escapable only by denial
of the whole enterprise.  Taken on their own terms they will, I believe, add
to our image of the world a certain coherency not present in existing
representations.  Whether they are accurate will be, of course, a matter for
investigation.

In this paper I define and exemplify conformally constrained metrics and
introduce the term \text{``space-time-\!-time''} in Sec\. II, exhibit canonical
decompositions of such metrics in Sec\. III, show in Sec\. IV how they
incorporate and relate metrical and electromagnetic gauge transformations, and
exhibit in Secs\. V and VI their connection forms and their geodesic equations
in frame systems adapted to the vector field $\xi$ of the constraint.  In
Sec\. VII, acting on the assumption that the geodesics of
\text{space-time-\!-time} describe histories of test particles, I define the
\text{space-time-\!-time} momentum covector of such a particle and use it to
make a physical interpretation of the \text{space-time-\!-time} geometry,
identifying certain scalars, vectors, and covectors along a geodesic as
electric charge, rest mass, space-time proper time, and space-time momentum of
the test particle in question, and certain geometrical fields as gravitational,
electromagnetic potential, electromagnetic bivector, and scalar gradient fields
exerting apparent forces on test particles in precisely determined ways.
Section VIII examines how the \text{space-time-\!-time} model distinguishes and
to what extent it relates the concepts of inertial mass, passive gravitational
mass, active gravitational mass, passive electric charge, and active electric
charge.  Section IX and the Appendix display the various curvature fields of a
conformally constrained metric:  curvature tensor, contracted curvature tensor,
curvature scalar, and Einstein tensor.  In Sec\. X I define and compute
``residual curvature'', an important concept peculiar to conformally
constrained metrics.  Lastly, Sec\. XI discusses the rationale for the term
\text{``space-time-\!-time''} and the need for extension of the conformally
constrained geometry that consistent application of that rationale implies.
\vskip 15pt

\noindent {\bf II.  Conformally Constrained Metrics}
\vskip 10pt

Let $\eusm M$ be a manifold and $\Ghat$ a (symmetric and nondegenerate) metric
on $\eusm M$.  That $\Ghat$ is {\bf conformally constrained} will mean that it
meets the following condition, in which ${\eusm L}_\xi$ denotes Lie
differentiation along $\xi$.
\vskip 5pt
\noindent {\bf Conformality Constraint.}  There exists on $\eusm M$ a vector
field $\xi$ such that ${\eusm L}_\xi \Ghat = 2 G$, where
$G := \Ghat - (\Ghat \xi \xi)^{-1} (\Ghat \xi \otimes \Ghat \xi)$.
\vskip 5pt

\noindent (The metric $\Ghat$ is understood to be a ``cocotensor'' field:  if
$P$ is a point of $\eusm M$, then $\Ghat (P)$ is an element of $\, T_P \otimes
T_P$, that is, a linear mapping of the tangent space $\, T^P$ of $\eusm M$ at
$P$ into $\, T_P$, the cotangent space of $\eusm M$ at $P$, regarded as the
dual space of $\, T^P$.  Thus $\Ghat \xi$ is a covector field on $\eusm M$,
and $\Ghat \xi \xi$ is a scalar field on $\eusm M$, the ``square length'' of
$\xi$ under $\Ghat$.  Implicit in the conformality constraint is that
$\Ghat \xi \xi$ vanishes nowhere, that, to put it differently, $\xi$ is
nowhere null with respect to $\Ghat$; a consequence is that $\xi$ itself
vanishes nowhere.  The symmetric cocotensor field $G$ is just the orthogonal
projection of $\Ghat$ along $\xi$, so the condition
${\eusm L}_\xi \Ghat = 2 G$ causes the lengths of vectors orthogonal to
and Lie transported by $\xi$ to expand.)

The prototype of conformally constrained metrics is the de Sitter space-time
metric, which in the Lema\^{\i}tre coordinate system takes the form    
$$
\Ghat = e^{2t} (dx \otimes dx + dy \otimes dy + dz \otimes dz)
                - R^2 (dt \otimes dt),
\tag1
$$
where $R$ is the (uniform) space-time radius of curvature \cite{1, 5}.  Here
$\xi = \partial / \partial t$, $\Ghat \xi = -R^2 dt$, $\Ghat \xi \xi = -R^2$,
and $G = e^{2t} (dx \otimes dx + dy \otimes dy + dz \otimes dz)$.  The manifold
$\eusm M$ covered by the Lema\^{\i}tre coordinate system is (together with
$\Ghat$) only half of the complete de Sitter space-time, which is a
single-sheeted hyperboloidal ``sphere'' $\eusm H$ of radius $R$ in the
Minkowski space $M(4,1)$.  Though not geodesically complete, $\eusm M$ is
$\xi$-{\bf complete} in that on every $\xi$-path (that is, on every maximally
extended integral path of $\xi$) the integration parameter runs from $-\infty$
to $\infty$.  Because $\eusm H$ is homogeneous, it is a union of open
``hemispheres'' like $\eusm M$, on each of which the metric of $\eusm H$ is
conformally constrained and $\xi$-complete.

Two additional examples of conformally constrained metrics are the
hyper-de Sitter metrics $\Ghat_\pm$ given by
$$
\Ghat_\pm = e^{2 \zeta} (dx \otimes dx + dy \otimes dy + dz \otimes dz
                         - dt \otimes dt) \pm R^2 (d\zeta \otimes d\zeta),
\tag2
$$
defined on manifolds $\eusm M_\pm$ that (with $\Ghat_\pm$) are open halves of
the two kinds of ``spheres'' of radius $R$ found in $M(4,2)$.  For both
metrics $\xi = \partial / \partial \zeta$ and $G =
e^{2 \zeta} (dx \otimes dx + dy \otimes dy + dz \otimes dz - dt \otimes dt)$;
but $\Ghat_+ \xi \xi = R^2$, whereas $\Ghat_- \xi \xi = -R^2$, which of course
reflects the fact that $\Ghat_+$ has diagonal signature $+++-+$ and $\Ghat_-$
has it $+++--$.  Both $\eusm M_+$ and $\eusm M_-$ are $\xi$-complete.

With these examples in mind let us agree to describe $\Ghat$ as
$\xi$-{\bf completely conformally constrained} if $\Ghat$ is conformally
constrained and $\eusm M$ is $\xi$-complete (with respect to the vector
field $\xi$ of the constraint), and as {\bf locally} ($\xi$-{\bf completely})
{\bf conformally constrained} if $\eusm M$ is a union of open submanifolds on
each of which the restriction of $\Ghat$ is ($\xi$-completely) conformally
constrained.  Then the metric of the de Sitter sphere $\eusm H$ is locally,
$\xi$-completely conformally constrained, as are the hyper-de Sitter sphere
metrics that extend $\Ghat_+$ and $\Ghat_-$.

The five-dimensional Kaluza metrics are characterized by the ``cylinder 
condition'' ${\eusm L}_\xi \Ghat = 0$ \cite{3}, which makes
$\xi$ a Killing vector field of, hence ``isometrically constrains'', $\Ghat$
\cite{6}.  Also, as readily follows, $\botsmash{{\eusm L}_\xi} G = 0$, so $G$
is Lie-constant along every $\xi$-path.  This projection $G$ of the metric
$\Ghat$, defined on the five-dimensional manifold of $\Ghat$, is essentially 
four-dimensional, being degenerate in the direction of $\xi$.  It was intended
(by Klein \cite{3a} and by Einstein \cite{3b}, each of whom adopted it in
preference to Kaluza's noninvariant alternative) to supplant the
four-dimensional metric of space-time, and was therefore supposed to have
diagonal signature $+++\,-$ for its nondegenerate part.  Having to choose
between $+++-+$ and $+++--$ for the signature of the full metric $\Ghat$,
Kaluza apparently opted for $+++-+$ \cite{7}.  As the first three +'s refer
to spatial dimensions, one naturally is tempted to say (and many do say) that
this causes Kaluza's extra dimension to be spatial also, and to call a Kaluza
manifold a \text{``space-time-\!-space''}.  But that is mere verbal analogy ---
it lacks any real justification in the form of a connection between the fifth
coordinate, generated along $\xi$, and the three dimensions of physical space
represented by the first three coordinates.

Rather than settle on one of these signatures for $\Ghat$, I shall proceed
as if either may be the case, and shall apply the descriptive term
\text{\bf space-time-\!-time} to every five-dimensional manifold $\eusm M$
bearing a locally, $\xi$-completely conformally constrained metric $\Ghat$, of
diagonal signature $+++-+$ (equivalently, $---+-$) or of signature $+++--$
(equivalently, $---++$), whose orthogonal projection along $\xi$ has a
space-time signature \cite{8}.  I intend in a subsequent paper to place the
two kinds of \text{space-time-\!-time} metric on equal footing as projections
of a single, higher dimensional, conformally constrained metric.  Physical
interpretations aside, all the computations that follow will be valid whatever
the dimensionality of $\eusm M$ or the signature of $\Ghat$.
\vskip 15pt

\noindent {\bf III.  Standard Forms of a Conformally Constrained Metric}
\vskip 10pt

Let $\Ghat$ be a metric that is conformally constrained, and
let $\eusm M$ be its carrying manifold.  One sees easily that
$$
\Ghat = G + \epshat \phi^2 (A \otimes A),
\tag3
$$
where $\phi := |\Ghat \xi \xi|^\frac12$, $A := (\Ghat \xi \xi)^{-1} \Ghat \xi$,
and $\epshat := 1$ if $\Ghat \xi \xi > 0$, but $-1$ if $\Ghat \xi \xi < 0$.
The projected metric $G$, the scalar field $\phi$, and the covector field $A$
behave in the following ways under Lie differentiation along $\xi$:
${\eusm L}_\xi \phi = 0$, ${\eusm L}_\xi A = 0$, and ${\eusm L}_\xi G = 2 G$.
This is demonstrable by a few simple calculations.  First, $G \xi =
\Ghat \xi - (\Ghat \xi \xi)^{-1} (\Ghat \xi \otimes \Ghat \xi) \xi =
\Ghat \xi - (\Ghat \xi \xi)^{-1} (\Ghat \xi \xi) \Ghat \xi = 0$.  Next, because
${\eusm L}_\xi \xi = 0$, one has that ${\eusm L}_\xi (\Ghat \xi) =
({\eusm L}_\xi \Ghat) \xi = 2 G \xi = 0$, and ${\eusm L}_\xi (\Ghat \xi \xi) =
({\eusm L}_\xi (\Ghat \xi)) \xi = 0$, so that clearly ${\eusm L}_\xi \phi = 0$
and ${\eusm L}_\xi A = 0$.  From Eq\. (3) it then follows that
${\eusm L}_\xi G = {\eusm L}_\xi \Ghat$, whence ${\eusm L}_\xi G = 2 G$.

A decomposition of $G$ comes about from solving the differential equation
${\eusm L}_\xi G = 2 G$.  Specifically, if $C$ is a scalar field on $\eusm M$,
then ${\eusm L}_\xi (e^{-2C} G) =
e^{-2C} [-2({\eusm L}_\xi C) G + {\eusm L}_\xi G] =
-2e^{-2C} ({\eusm L}_\xi C - 1) G$.  If ${\eusm L}_\xi C = 1$, then
${\eusm L}_\xi \Go = 0$, where $\Go := e^{-2C} G$.  Thus $G = e^{2C} \Go$, and
$$
\Ghat = e^{2C} \Go + \epshat \phi^2 (A \otimes A),
\tag4
$$
where $C$ is a scalar field, ${\eusm L}_\xi C = 1$, $\Go$ is a metric on
$\eusm M$ of the same signature and degeneracy as $G$, and
${\eusm L}_\xi \Go = 0$.  Application of ${\eusm L}_\xi$ to both sides of
Eq\. (4) shows that this representation for $\Ghat$, under the conditions that
${\eusm L}_\xi C = 1$ and the Lie derivatives along $\xi$ of $\Go$, $\phi$,
and $A$ all vanish, is sufficient to make $\Ghat$ satisfy the conformality
constraint (with respect to $\xi$).  With these conditions the representation
therefore constitutes a characterization of conformally constrained metrics.

Continuing, let us introduce (by a standard construction) a coordinate
system $\[[ x^\mu, \zeta \]]$ adapted to $\xi$ so that $\xi =
\partial / \partial \zeta$.  (Here $\mu$ and other Greek letter indices will
range, if $d > 1$, from 1 to $d - 1$, where $d := \text{dim} \; {\eusm M}$;
if $d = 1$, then the only coordinate is $\zeta$, so $\mu$ does not enter the
game.)  As a covector field, $A$ has in $\[[ x^\mu, \zeta \]]$ the expansion
$A = A_\mu dx^\mu + A_\zeta d\zeta$.  But $A_\zeta =
A(\partial / \partial \zeta) = A \xi = (\Ghat \xi \xi)^{-1} \Ghat \xi \xi = 1$,
so $A = A_\mu dx^\mu + d\zeta$.  Further, $0 = {\eusm L}_\xi A =
{\eusm L}_{\partial / \partial \zeta} A =
(\partial A_\mu / \partial \zeta) dx^\mu$, so
$\partial A_\mu / \partial \zeta = 0$; thus the $A_\mu$ depend on the
coordinates $x^\kappa$ alone, and not on $\zeta$.  Also,
$\partial \phi / \partial \zeta = {\eusm L}_\xi \phi = 0$, so $\phi$ is a
function of the $x^\kappa$ only.  The projected metric $G$ has the expansion
$G = dx^\mu \otimes g_{\mu \nu} dx^\nu + dx^\mu \otimes g_{\mu \zeta} d\zeta +
d\zeta \otimes g_{\zeta \nu} dx^\nu + d\zeta \otimes g_{\zeta \zeta} d\zeta$.
But $0 = G \xi = G(\partial / \partial \zeta) =
g_{\zeta \nu} dx^\nu + g_{\zeta \zeta} d\zeta$, so
$g_{\zeta \nu} = g_{\zeta \zeta} = 0$.  Because $G$ is symmetric,
$g_{\mu \zeta}$ vanishes also, and therefore
$G = dx^\mu \otimes g_{\mu \nu} dx^\nu$.  The condition
${\eusm L}_\xi G = 2 G$ translates to
$\partial g_{\mu \nu} / \partial \zeta = 2 g_{\mu \nu}$.
In this way we arrive at the adapted coordinates version of Eq\. (3), viz.
$$
\Ghat = dx^\mu \otimes g_{\mu \nu} dx^\nu
         + \epshat \phi^2 (A_\mu dx^\mu + d\zeta) \otimes
                          (A_\nu dx^\nu + d\zeta),
\tag"($3{\bold '}$)"
$$
with $\partial \phi / \partial \zeta = \partial A_\mu / \partial \zeta = 0$
and $\partial g_{\mu \nu} / \partial \zeta = 2 g_{\mu \nu}$.

To do the same for Eq\. (4), let us now select the scalar field $C$.  Because
$\partial C / \partial \zeta = {\eusm L}_\xi C = 1$, the possibilities are
$C = \zeta + \theta$, thus $e^{2 C} = e^{2 \zeta} e^{2 \theta}$, with
$\partial \theta / \partial \zeta = 0$.  But the factor $e^{2 \theta}$ can be
absorbed by redefining $\Go$, so let us take $C = \zeta$.  Then
$\Go = e^{-2 \zeta} G = dx^\mu \otimes \go_{\mu \nu} dx^\nu$,
where $\go_{\mu \nu} := e^{-2 \zeta} g_{\mu \nu}$;
consequently, $G = e^{2 \zeta} \Go$,
$g_{\mu \nu} = e^{2 \zeta} \go_{\mu \nu}$, and, because
${\eusm L}_\xi \Go = 0$, $\partial \go_{\mu \nu} / \partial \zeta = 0$.
Let us also introduce the covector field $\Ao := A - d\zeta$, for
which $A = \Ao + d\zeta$, $\Ao = \Ao_\mu dx^\mu$, $\Ao_\mu = A_\mu$,
${\eusm L}_\xi \Ao = {\eusm L}_\xi A = 0$, and
$\partial \Ao_\mu / \partial \zeta = \partial A_\mu / \partial \zeta = 0$.
Then Eq\. (4) takes the forms
\vskip -8pt
$$
\aligned
\Ghat
 &= e^{2 \zeta} \Go
      + \epshat \phi^2 \L(\Ao + d \zeta\R) \otimes \L(\Ao + d \zeta\R) \\
 &= e^{2 \zeta} (dx^\mu \otimes \go_{\mu \nu} dx^\nu)
      + \epshat \phi^2 \L(\Ao_\mu dx^\mu + d\zeta\R) \otimes
                       \L(\Ao_\nu dx^\nu + d\zeta\R),
\endaligned \tag"($4{\bold '}$)"
$$
with $\partial \phi / \partial \zeta = \partial \Ao_\mu / \partial \zeta =
\partial \go_{\mu \nu} / \partial \zeta = 0$.

The ability of a metric $\Ghat$ to assume the standard forms $(3')$ and $(4')$
with the stated conditions on $\phi$, $A_\mu$, $\Ao_\mu$, $g_{\mu \nu}$, and
$\go_{\mu \nu}$ satisfied is necessary and sufficient for the restrictions of
$\Ghat$ to the domains of all such adapted coordinate systems
$\[[ x^\mu, \zeta \]]$ to be conformally constrained, thus for $\Ghat$ to be
locally conformally constrained.  
\vskip 15pt

\noindent {\bf IV.  Gauge Transformations}
\vskip 10pt

Coordinate systems adapted to $\xi$ such as $\[[ x^\mu, \zeta \]]$ of the
preceding section may be constructed in the following well-known way.
Pick a hypersurface $\eusm S$ of $\eusm M$ that is transverse to $\xi$, and a
coordinate system $\[[ y^\mu \]]$ of $\eusm S$, and suppose that no $\xi$-path
crosses $\dom \[[ y^\mu \]]$ twice \cite{9}.  For each point $P$ of
$\eusm M$ that lies on a trajectory of $\xi$ (that is, in some $\xi$-path's
range) whose intersection with $\eusm S$ is a point $Q$ in the domain of
$\[[ y^\mu \]]$, let $x^\mu (P) = y^\mu (Q)$ and let $\zeta (P)$ be the value
attained at $P$ by the integration parameter of $\xi$ that starts with the
value 0 at $Q$.  Then $\[[ x^\mu, \zeta \]]$ is a coordinate system of
$\eusm M$ whose domain is the set of all such points $P$.  It is adapted to
$\xi$ in the sense that $\xi = \partial / \partial \zeta$, and to $\eusm S$
in that $\zeta |_{\eusm S} = 0$.

The only arbitrary elements in this construction are the hypersurface $\eusm S$
and the coordinate system $\[[ y^\mu \]]$ of $\eusm S$.  When one picks a
different hypersurface ${\eusm S}'$ transverse to $\xi$, and $\xi$-transfers
$\[[ y^\mu \]]$ to ${\eusm S}'$ to use as the coordinate system
$\[[ y^{\mu'} \]]$ of ${\eusm S}'$, so that $y^{\mu'} (Q') = y^\mu (Q)$ if
$Q$ in $\eusm S$ and $Q'$ in ${\eusm S}'$ belong to the same trajectory of
$\xi$, then the coordinate system $\[[ x^{\mu'}, \zeta' \]]$ produced by the
construction is related to $\[[ x^\mu, \zeta \]]$ by $x^{\mu'} = x^\mu$ and
$\zeta' = \zeta - \lambda$, where $\lambda := \zeta - \zeta'$.  The scalar
field $\lambda$ is constant on each trajectory of $\xi$ traversing its domain,
hence is independent of $\zeta$, for if $Q$ and $Q'$ are the points where the
$\xi$-trajectory intersects $\eusm S$ and ${\eusm S}'$, respectively, then
$\lambda (P) = \zeta (Q') = -\zeta' (Q)$ for every point $P$ on the trajectory.

From $\zeta' = \zeta - \lambda$ it follows that $d\zeta = d\lambda + d\zeta' =
(\partial \lambda / \partial x^\mu) dx^\mu + d\zeta'$, hence that the covector
field $A$, which has in $\[[ x^\mu , \zeta \]]$ the expansion
$A = \Ao + d\zeta = \Ao_\mu dx^\mu + d\zeta$, has in $\[[ x^\mu, \zeta' \]]$
the expansion $A = \Ao' + d\zeta' = \Ao'_\mu dx^\mu + d\zeta'$ with
$\Ao' = \Ao + d\lambda$ and, consequently,
$\Ao_\mu + \partial \lambda / \partial x^\mu$.  In the event that $\Ghat$
is a \text{space-time-\!-time} metric, the negative of twice the exterior
differential of $A$ will come to be identified as the electromagnetic field
tensor $F$.  We shall have then that
$F = -2d_\wedge A = -2d_\wedge \L(\Ao + d\zeta\R) = -2d_\wedge \Ao$,
hence that $\Ao$ plays the role of electromagnetic four-vector potential.
But we shall have also that $F = -2d_\wedge \Ao'$, so that $\Ao'$ plays the
same role, but in a different gauge.  This tells us that the transformation
from the adapted coordinate system $\[[ x^\mu, \zeta \]]$ to the adapted
coordinate system $\[[ x^\mu, \zeta' \]]$ generates a gauge transformation
$\Ao \to \Ao + d\lambda$ of the electromagnetic four-vector potential.  The
converse likewise is true:  every gauge transformation $\Ao \to \Ao + d\lambda$
with $\lambda$ a scalar field independent of $\zeta$ determines a
transformation from the adapted coordinate system $\[[ x^\mu, \zeta \]]$ to an
adapted coordinate system $\[[ x^\mu, \zeta' \]]$ with
$\zeta' = \zeta - \lambda$.

The discussion up to this point only recapitulates what Klein \cite{3a} and
Einstein \cite{3b} worked out long ago for the Kaluza (--Klein) geometry.
Their identification of electromagnetic four-potential gauge transformations
with adapted-coordinates transformations in five dimensions was the first step
on the road to the gauge theories that currently permeate theoretical physics.
Missing from Kaluza--Klein theory and from these gauge theories, however, is
any remembrance of Weyl's earlier association of electromagnetic gauge
transformations with (conformal) gauge transformations of the metric of
space-time \cite{10}.  In \text{space-time-\!-time} this association is
preserved, as we are now in position to see.

It is really quite simple.  When $\Ghat$ is a \text{space-time-\!-time} metric,
it is $\Go$ that takes the role of space-time metric.  But there is not just
one $\Go$, there are many, each corresponding to a particular choice of
the hypersurface $\eusm S$ in the construction of the adapted coordinates.
If, as before, $\eusm S$ and ${\eusm S}'$ are two such choices, then
$G = e^{2 \zeta} \Go = e^{2 \zeta'} \Go'$, where
$\Go' = e^{2 \lambda} \Go$.  Thus the same coordinate transformation that
generates the electromagnetic gauge transformation $\Ao \to \Ao + d\lambda$
generates Weyl's metrical gauge transformation $\Go \to e^{2 \lambda} \Go$.

The coordinate transformations that generate the electromagnetic and the
metrical gauge transformations, {\it being} coordinate transformations,
alter only the representation of the \text{space-time-\!-time} metric, not the
metric itself.  This is a principal
advantage that the \text{space-time-\!-time} geometry has over the Weyl
geometry.  Weyl, working before Kaluza first proposed using five dimensions to
unify gravity and electromagnetism, impressed his infinitude of conformally
related space-time metrics onto one four-dimensional manifold.  That is very
much like drawing all the maps of the world on a single sheet of paper, a
practice that would conserve paper but confound navigators.  In effect, the
\text{space-time-\!-time} geometry economizes on paper but avoids the confusion
of maps on maps, by drawing a selection of the maps on individual sheets, then
stacking the sheets so that each of the remaining maps can be generated on
command by slicing through the stack in a particular way.  Nothing is lost
thereby, and much is gained, as we shall see.
\vskip 15pt

\noindent {\bf V.  Connection Forms and Covariant Differentiations}
\vskip 10pt

Further study of the geometry of the conformally constrained metric $\Ghat$
will be facilitated if we work in a frame  system that on the one hand takes
full advantage of the orthogonality between $G$ and $\Ghat - G$ and on the
other hand is Lie constant along $\xi$, but that elsewise is unrestricted.
To accomplish this, let us back up a little and relabel the coordinate system
$\[[ x^\mu, \zeta \]]$ adapted to $\xi$ as $\[[ x^{\mu'}, \zeta \]]$.  Then
$A = A_{\mu'} dx^{\mu'} + \, d\zeta$,
$G = dx^{\mu'} \lotimes g_{\mu' \nu'} dx^{\nu'}$,
$\Ao = \Ao_{\mu'} dx^{\mu'}$, and
$\Go = dx^{\mu'} \lotimes \go_{\mu' \nu'} dx^{\nu'}$.  Now let
$\omega^d := \phi A$ (recall that $d := \text{dim} \; {\eusm M}$),
and let $\{ \omega^\mu \}$ be any pointwise linearly independent ordered set
of $d - 1$ covector fields that are smooth linear combinations of the
$dx^{\mu'}$ with coefficients independent of $\zeta$.  Then
$\omega^\mu = dx^{\mu'} J_{\mu'}{}^\mu$ and
$dx^{\mu'} = \omega^\mu J_\mu{}^{\mu'}$, where
$[J_{\mu'}{}^\mu]$ and $[J_\mu{}^{\mu'}]$ are reciprocal matrix fields and
satisfy $\partial J_{\mu'}{}^\mu / \partial \zeta =
\partial J_\mu{}^{\mu'} / \partial \zeta = 0$.
The ordered set $\{ \omega^\mu, \omega^d \}$ is also pointwise linearly
independent; it therefore is a coframe system of $\eusm M$, defined on the
domain of the coordinate system $\[[ x^{\mu'}, \zeta \]]$.  In this coframe
system one has that $A = \phi^{-1} \omega^d$,
$G = \omega^\mu \otimes g_{\mu \nu} \omega^\nu$, and
$\Go = \omega^\mu \otimes \go_{\mu \nu} \omega^\nu$, hence that
$$
\Ghat = \omega^\mu \otimes g_{\mu \nu} \omega^\nu
         + \epshat (\omega^d \otimes \omega^d)
\tag"($3{\bold {''}}$)"
$$
and
$$
\Ghat = e^{2 \zeta} (\omega^\mu \otimes \go_{\mu \nu} \omega^\nu)
         + \epshat (\omega^d \otimes \omega^d),
\tag"($4{\bold {''}}$)"
$$
where $g_{\mu \nu} = J_\mu{}^{\mu'} g_{\mu' \nu'} J_\nu{}^{\nu'}$ and
$\go_{\mu \nu} = J_\mu{}^{\mu'} \go_{\mu' \nu'} J_\nu{}^{\nu'}$, with the
consequences that $\partial g _{\mu \nu} / \partial \zeta = 2 g_{\mu \nu}$
and $\partial \go _{\mu \nu} / \partial \zeta = 0$.  Also,
$\Ao = \Ao_\mu \omega^\mu$, where $\Ao_\mu = J_\mu{}^{\mu'} \Ao_{\mu'}$, 
$\Ao_{\mu'} = A_{\mu'}$, and, consequently,
$\partial \Ao_\mu / \partial \zeta = 0$ (note that in general
$\Ao_\mu \neq A_\mu$, even though $A$ has the mixed expansion
$A = \Ao_\mu \omega^\mu + d\zeta$; in fact $A_\mu = 0$ and $A_d = \phi^{-1}$).

Upon identifying the frame system $\{ e_\mu, e_d \}$ to which
$\{ \omega^\mu, \omega^d \}$ is dual, one has
$$
\align
e_\mu
 &= J_\mu{}^{\mu'} (\partial / \partial x^{\mu'})
         - \Ao_\mu (\partial / \partial \zeta) \quad \text{and} \quad
    e_d = \phi^{-1} \xi = \phi^{-1} (\partial / \partial \zeta),
\tag5 \\
\intertext{to go with}
\omega^\mu 
 &=dx^{\mu'} J_{\mu'}{}^\mu \quad \text{and} \quad
    \omega^d = \phi A = \phi (A_{\mu'} dx^{\mu'} + d\zeta)
                      = \phi (\Ao_\mu \omega^\mu + d\zeta).
\tag6
\endalign
$$
The vector field $e_d$ is the unit normalization of $\xi$ and is orthogonal
to each of the vector fields $e_\mu$.  It is not difficult to see that
${\eusm L}_\xi e_\mu = {\eusm L}_\xi e_d = 0$ and
${\eusm L}_\xi \omega^\mu = {\eusm L}_\xi \omega^d = 0$.  Thus we have a frame
system and its dual coframe system that are Lie constant along $\xi$, but with
the further property that $e_d$ has length 1 and is orthogonal to each $e_\mu$.
Their constancy along $\xi$ makes them gauge invariant: the adapted coordinates
transformation $\[[ x^{\mu'}, \zeta \]] \to \[[ x^{\mu'}, \zeta' \]]$ has no
effect on them.  This is what we sought.  Borrowing terminology from fibre 
bundle theory we may call the $e_\mu$ and the tangent subspace they span at a
point ``horizontal'', and $e_d$ and the subspace it spans at a point
``vertical'', as determined with reference to the covector field $A$, standing
in for a bundle connection 1-form.

To identify differentiations of a scalar field $f$ by the various frame
operators, let us adopt the abbreviations
$f_{. \mu'} := \partial f / \partial x^{\mu'}$,
$f_{. \zeta} := \partial f / \partial \zeta$,
$f_{, \mu} := e_\mu f$, and $f_{, d} := e_d f$; also,
let $f_{. \mu} := J_\mu{}^{\mu'} f_{. \mu'}$.  Then from Eqs\. (5) follow
$$
f_{, \mu} = J_\mu{}^{\mu'} f_{. \mu'} - \Ao_\mu f_{. \zeta} =
f_{. \mu} - \Ao_\mu f_{. \zeta} \quad \text{and} \quad
f_{, d} = \phi^{-1} f_{. \zeta}.
\tag7
$$
When $f$ is independent of $\zeta$, then $f_{, d} = 0$ and
$f_{, \mu} = f_{. \mu}$.  In particular
$$
\phi_{. \zeta} = \Ao_{\mu . \zeta} = \go_{\mu \nu . \zeta}
               = J_{\mu'}{}^\mu{}_{. \zeta} = J_\mu{}^{\mu'}\!{}_{. \zeta} = 0,
\tag8
$$
so
$$
\Ao_{\mu , d} = \go_{\mu \nu , d} = J_{\mu'}{}^\mu{}_{, d}
              = J_\mu{}^{\mu'}\!{}_{, d} = 0
\tag9
$$
and
$$
\phi_{, \mu} = \phi_{. \mu}, \quad \Ao_{\mu , \nu} = \Ao_{\mu . \nu},
\quad \go_{\mu \nu , \kappa} = \go_{\mu \nu . \kappa},
$$
$$
J_{\mu'}{}^\mu{}_{, \nu} = J_{\mu'}{}^\mu{}_{. \nu}, \quad \text{and} \quad
J_\mu{}^{\mu'}\!{}_{, \nu} = J_\mu{}^{\mu'}\!{}_{. \nu}.
\tag10
$$
On the other hand,
$$
\align
g_{\mu \nu . \zeta}
 &= 2 g_{\mu \nu},
\tag11 \\
\intertext{so}
g_{\mu \nu , d}
 &= 2 \phi^{-1} g_{\mu \nu}
\tag12 \\
\intertext{and}
g_{\mu \nu , \kappa}
 &= g_{\mu \nu . \kappa} - 2 g_{\mu \nu} \Ao_\kappa.
\tag13
\endalign
$$

For the exterior differential of $\omega^\mu$ we have $d_\wedge \omega^\mu =
C_\kappa{}^\mu{}_\lambda \omega^\lambda \lwedge \omega^\kappa$,
with $C_\kappa{}^\mu{}_\lambda$ skew-symmetric in $\kappa$ and $\lambda$ and
independent of $\zeta$, so that
$$
C_\kappa{}^\mu{}_{\lambda . \zeta} = 0, \quad
    C_\kappa{}^\mu{}_{\lambda , d} = 0, \quad \text{and} \quad
  C_\kappa{}^\mu{}_{\lambda , \nu} = C_\kappa{}^\mu{}_{\lambda . \nu}
\tag14
$$
(in terms of $J_{\mu'}{}^\mu$ and $J_\mu{}^{\mu'}$,
$C_\kappa{}^\mu{}_\lambda = J_{[\kappa}{}^{\mu'} J_{\mu'}{}^\mu{}_{. \lambda]}$
\cite{11}).  From $\omega^d = \phi A$ it follows that
$d_\wedge \omega^d = -(1/2) \phi F + d\phi \wedge A$, where
$$
\align
 &\aligned
  F :\!@!@!@!@!@!@!@!
   &= -2 d_\wedge A = -2 d_\wedge \Ao \\ 
   &= F_{\kappa \lambda} \omega^\lambda \lwedge \omega^\kappa
    = F_{\kappa \lambda} \omega^\lambda \otimes \omega^\kappa
  \endaligned
\tag15 \\
\intertext{with}
 &\hskip -4pt
  \aligned
  F_{\kappa \lambda}
   &= -2 \L(\Ao_{[\kappa . \lambda]}
               + \Ao_\mu C_\kappa{}^\mu{}_\lambda\R) \\
   &= \Ao_{\lambda . \kappa} - \Ao_{\kappa . \lambda}
       - 2 \Ao_\mu C_\kappa{}^\mu{}_\lambda,
  \endaligned
\tag16
\endalign
$$
in consequence of which
$$
F_{\kappa \lambda . \zeta} = 0, \quad
    F_{\kappa \lambda , d} = 0, \quad \text{and} \quad
  F_{\kappa \lambda , \mu} = F_{\kappa \lambda . \mu}.
\tag17
$$
Thus
$$
\align
d_\wedge \omega^\mu
 &= C_\kappa{}^\mu{}_\lambda \omega^\lambda \lwedge \omega^\kappa,
\tag18 \\
\intertext{and}
d_\wedge \omega^d
 &= -(1/2) \phi F_{\kappa \lambda} \omega^\lambda \lwedge \omega^\kappa
      + \phi^{-1} \phi_{. \lambda} \omega^\lambda \lwedge \omega^d.
\tag19
\endalign
$$
By use of Eqs\. (18) and (19), and the fact that, for
$K, L, M = 1, \ldots@! ,d$,
$[e_K, e_L] = C_K{}^M\!{}_L \, e_M$ if $d_\wedge \omega^{M}
= C_K{}^M\!{}_L \, \omega^L \lotimes \omega^K$, the nonvanishing commutators of
the frame system $\{ e_\mu, e_d \}$ are readily expressed \cite{12}:
$$
\align
[e_\kappa, e_\lambda]
 &= C_\kappa{}^\mu{}_\lambda e_\mu - (1/2) \phi F_{\kappa \lambda} e_d
\tag20 \\
\intertext{and}
[e_\kappa, e_d]
 &= -(1/2) \phi^{-1} \phi_{. \kappa} e_d = -[e_d, e_\kappa].
\tag21
\endalign
$$

Let us denote by $\bdhat$ the torsionless covariant differentiation 
on $\eusm M$ that is compatible with $\Ghat$, and by
$\omegahat_\kappa{}^\mu$, $\omegahat_\kappa{}^d$, $\omegahat_d{}^\mu$,
and $\omegahat_d{}^d$ the connection forms of $\bdhat$ in the frame system
$\{e_\mu, e_d \}$, so that $\bdhat \Ghat = 0$ and
$$
\align
\bdhat e_\kappa
 &= \omegahat_\kappa{}^\mu \otimes e_\mu + \omegahat_\kappa{}^d \otimes e_d, \\
\bdhat e_d
 &= \omegahat_d{}^\mu \otimes e_\mu + \omegahat_d{}^d \otimes e_d, \\
\vspace{-7pt}
\tag22 \\
\vspace{-7pt}
\bdhat \omega^\mu
 &= -\omegahat_\kappa{}^\mu \otimes \omega^\kappa
      - \omegahat_d{}^\mu \otimes \omega^d, \\
\vspace{-14pt}
\intertext{and}
\vspace{-14pt}
\bdhat \omega^d
 &= -\omegahat_\kappa{}^d \otimes \omega^\kappa
      - \omegahat_d{}^d \otimes \omega^d.
\endalign
$$
By standard methods these connection forms can be expressed in terms of the
metric components in Eq\. ($3''$) and the exterior differential coefficients
in Eqs\. (18) and (19).  The result is that
$$
\align
\omegahat_\kappa{}^\mu
 &= \omega_\kappa{}^\mu
     + \L[\phi^{-1} g_\kappa{}^\mu
             + \epshat (1/2) \phi F_\kappa{}^\mu\R] \omega^d, \\
\omegahat_\kappa{}^d
 &= -\L[\epshat \phi^{-1} g_{\kappa \lambda}
           - (1/2) \phi F_{\kappa \lambda}\R] \omega^\lambda
      + \phi^{-1} \phi_{. \kappa} \omega^d, \\
\omegahat_d{}^\mu
 &= \L[\phi^{-1} g^\mu{}_\lambda
          - \epshat (1/2) \phi F^\mu{}_\lambda\R] \omega^\lambda
     - \epshat \phi^{-1} \phi^{. \mu} \omega^d
\tag23 \\ 
 &= -\epshat g^{\mu \kappa} \omegahat_\kappa{}^d, \\
\vspace{-14pt}
\intertext{and}
\vspace{-14pt}
\omegahat_d {}^d
 &= 0;
\endalign
$$
in these equations
$$
\aligned
\omega_\kappa{}^\mu
 &:= \Gamma_\kappa{}^\mu{}_\lambda \omega^\lambda, \\
\Gamma_\kappa{}^\mu{}_\lambda
 &:= \{ {}_\kappa{}^\mu{}_\lambda \}
      - (C_\kappa{}^\mu{}_\lambda
         + C_{\kappa \lambda}{}^\mu
         + C_{\lambda \kappa}{}^\mu), \\
\{ {}_\kappa{}^\mu{}_\lambda \}
 &:= (1/2) (g_{\nu \lambda , \kappa}
            + g_{\kappa \nu , \lambda}
            - g_{\kappa \lambda , \nu}) g^{\nu \mu}, \\
C_{\kappa \lambda}{}^\mu
 &:= g_{\lambda \nu} C_\kappa{}^\nu{}_\pi g^{\pi \mu}, \quad
     \phi^{. \mu} := \phi_{. \lambda} g^{\lambda \mu}, \\
F_\kappa{}^\mu
 &:= F_{\kappa \lambda} g^{\mu \lambda}, \quad 
     F^\mu{}_\lambda := g^{\kappa \mu} F_{\kappa \lambda}, \\
g_\kappa{}^\mu
 &:= g_{\kappa \lambda} g^{\lambda \mu} = \delta_\kappa{}^\mu,
     \quad \text{and} \quad
     g^\mu{}_\lambda := g^{\mu \nu} g_{\nu \lambda} = \delta^\mu{}_\lambda,
\endaligned
\tag24
$$
$[g^{\nu \mu}]$ being the matrix field inverse to $[g_{\mu \nu}]$.

An alternate covariant differentiation $\bold d$ on $\eusm M$ is fixed by 
the stipulations that $\bold d e_\kappa = \omega_\kappa{}^\mu \otimes e_\mu$
and $\bold d e_d = 0$, or, equivalently, that
$\bold d \omega^\mu = - \omega_\kappa{}^\mu \otimes \omega^\kappa$ and
$\bold d \omega^d = 0$.  It has the properties
i) $\bold d G = 2 A \otimes G$,
ii) $\text{Tor\ } \bold d = d_\wedge \omega^d \otimes e_d =
(d_\wedge A + \phi^{-1} d\phi \wedge A) \otimes \xi =
[-(1/2) F + \phi^{-1} d\phi \wedge A] \otimes \xi$, and
iii) $\bold d A = -\phi^{-1} d\phi \otimes A$.  Because $G$ is degenerate,
properties (i) and (ii) do not alone determine $\bold d$; but properties (i),
(ii), and (iii) do.  These properties are gauge invariant, and so, therefore,
is $\bold d$.  Property (iii), a reformulation of $\bold d \omega^d = 0$,
implies that $\bold d \Ghat = \bold d G$, hence that
$\bold d \Ghat = 2 A \otimes G$, in light of property (i).  Although $G$
has no inverse, it is useful to let
$G^{-1} := e_\mu \otimes g^{\mu \nu} e_\nu$, and then one sees that
$G^{-1} \Ghat = G^{-1} G = \omega^\mu \otimes e_\mu$,
$\Ghat G^{-1} = G G^{-1} = e_\mu \otimes \omega^\mu$, and
$\bold d \Ghat^{-1} = \bold d G^{-1} = -2 A \otimes G^{-1}$.
All connection forms and coefficients of $\bold d$ other than the
$\omega_\kappa{}^\mu$ and the $\Gamma_\kappa{}^\mu{}_\lambda$, that is, all
with $d$ as a suffix, vanish.  This covariant differentiation is an analog in
the conformally constrained geometry of the covariant differentiation (affine
connection) in Weyl's geometry, the principal characteristic of which is that
it satisfies the equation in property (i) above, properly interpreted.

Bringing into play Eq\. (13) we can express the Christoffel symbols
$\{ {}_\kappa{}^\mu{}_\lambda \}$ in the more expanded form
$$
\align
\{ {}_\kappa{}^\mu{}_\lambda \}
 &= (1/2) (g_{\nu \lambda . \kappa}
           + g_{\kappa \nu . \lambda}
           - g_{\kappa \lambda . \nu}) g^{\nu \mu} \\
 &\qquad - \L(g_{\nu \lambda} \Ao_\kappa
                + g_{\kappa \nu} \Ao_\lambda
                - g_{\kappa \lambda} \Ao_\nu\R) g^{\nu \mu}.
\tag25
\endalign
$$
A further breaking out arises from replacing $g_{\mu \nu}$ by
$e^{2 \zeta} \go_{\mu \nu}$ and, accordingly, $g^{\nu \mu}$ by
$e^{-2 \zeta} \go^{\nu \mu}$, where $[\go^{\nu \mu}]$ is the inverse of
$[\go_{\mu \nu}]$.  That results in
$$
\{ {}_\kappa{}^\mu{}_\lambda \}
  = \{ {}_\kappa{}^\mu{}_\lambda \rbraceo + \Deltao_\kappa{}^\mu{}_\lambda,
\tag26
$$
in which
$$
\aligned
\{ {}_\kappa{}^\mu{}_\lambda \rbraceo
 &:= (1/2) (\go_{\nu \lambda . \kappa}
            + \go_{\kappa \nu . \lambda}
            - \go_{\kappa \lambda . \nu}) \go^{\nu \mu}, \\
\Deltao_\kappa{}^\mu{}_\lambda
 &:= -\L(\Ao_\kappa \go^\mu{}_\lambda
                + \go_\kappa{}^\mu \Ao_\lambda
                + \go_{\kappa\lambda} \Ao^\mu\R), \\
\Ao^\mu
 &:= \Ao_\lambda \go^{\lambda \mu}, \\
\go^\mu{}_\lambda
 &:= \go^{\mu \nu} \go_{\nu \lambda} = \delta^\mu{}_\lambda,
     \quad \text{and} \quad
     \go_\kappa{}^\mu := \go_{\kappa \lambda} \go^{\lambda \mu}
                       = \delta_\kappa{}^\mu.
\endaligned \tag27
$$
This in turn gives
$$
\align
\omega_\kappa{}^\mu
 &= \omegao_\kappa{}^\mu + \Deltao_\kappa{}^\mu \\
\vspace{-14pt}
\intertext{and}
\vspace{-14pt}
\vspace{-14pt}
\tag28 \\
\Gamma_\kappa{}^\mu{}_\lambda
 &= \Gammao_\kappa{}^\mu{}_\lambda + \Deltao_\kappa{}^\mu{}_\lambda,
\endalign
$$
where
$$
\align
\omegao_\kappa{}^\mu
 &:= \Gammao_\kappa{}^\mu{}_\lambda \omega^\lambda, \quad
     \Deltao_\kappa{}^\mu := \Deltao_\kappa{}^\mu{}_\lambda \omega^\lambda,\\
\Gammao_\kappa{}^\mu{}_\lambda
 &:= \{ {}_\kappa{}^\mu{}_\lambda \rbraceo
       - \L(C_\kappa{}^\mu{}_\lambda
               + \Co_{\kappa \lambda}{}^\mu
               + \Co_{\lambda \kappa}{}^\mu\R),
\tag29 \\
\vspace{-14pt}
\intertext{and}
\vspace{-14pt}
\Co_{\kappa \lambda}{}^\mu
 &:= \go_{\lambda \nu} C_\kappa{}^\nu{}_\pi \go^{\pi \mu} \quad
     (= C_{\kappa \lambda}{}^\mu, \text{\ as well)}.
\endalign
$$

Yet another covariant differentiation $\bdo$ on $\eusm M$ is fixed by the
stipulations that $\bdo {e_\kappa} = \omegao_\kappa{}^\mu \otimes e_\mu$ and
$\bdo {e_d} = 0$, which are equivalent to $\bdo {\omega^\mu} =
- \omegao_\kappa{}^\mu \otimes \omega^\kappa$ and $\bdo {\omega^d} = 0$.
It possesses and is determined by the properties i) $\bdo \Go = 0$,
ii) $\text{Tor\ } \bdo = \text{Tor\ } \bold d$, and iii) $\bdo A = \bold d A$,
but like $\bold d$ it is not determined by (i) and (ii) alone.
If $\Go^{-1} := e_\mu \otimes \go^{\mu \nu} e_\nu$, then
$\Go^{-1} \Go = \omega^\mu \otimes e_\mu$,
$\Go \Go^{-1} = e_\mu \otimes \omega^\mu$, and $\bdo \Go^{-1} = 0$.  All
connection forms and coefficients of $\bdo$ other than the
$\omegao_\kappa{}^\mu$ and the $\Gammao_\kappa{}^\mu{}_\lambda$ vanish.  Unlike
$\bold d$, which, being determined by gauge invariant properties, is itself
gauge invariant, $\bdo$ is not gauge invariant.  That is to say, each new
choice of a gauge brings with it a new $\Go$, and with that comes a (usually)
new $\bdo$ compatible with the new $\Go$.  This covariant differentiation is,
in the \text{space-time-\!-time} case, a generalized analog of the usual
space-time covariant differentiation.

The formulas displayed above will enable us to write out in reasonably 
comprehensible form the geodesic equations and the various curvature tensor
fields of the conformally constrained geometry.  Some of their terms disappear
in the corresponding formulas for the Kaluza geometry, which is described by
the metric of Eq\. ($4'$) with the factor $e^{2 \zeta}$ removed; in the
Kaluza--Klein geometry, which has in addition $\phi$ = constant, the terms
involving derivatives of $\phi$ disappear as well.  Thus in the conformally
constrained geometry there are more hooks to hang physical interpretations on
than in the Kaluza geometry, and even more yet than in the Kaluza--Klein
geometry.

One aspect of the Kaluza and the Kaluza--Klein geometries that persists
in the conformally constrained geometry is that the vanishing of
the 2-form $F$ is necessary (and sufficient) for the possibility of gauging
away to zero the potential field $\Ao$.  Specifically, if $F = 0$, then
$d_\wedge \Ao = 0$, so (locally) there exists a scalar field $\lambda$ such
that $\Ao = -d\lambda$, hence such that
$\Ao_\mu \omega^\mu = - \lambda_{, \mu} \omega^\mu - \lambda_{, d} \omega^d$.
But then $\lambda_{, d} = 0$, so $\lambda_{. \zeta} = 0$, and if
$\zeta' = \zeta - \lambda$, then $\Ao' = \Ao + d\lambda = 0$.  An important
distinction, however, is that, whereas in the Kaluza and the Kaluza--Klein
geometries $\Ao$ may be thus gauged away without disturbing the metric
$\Go$, in the conformally constrained geometry the gauging away of $\Ao$ is
inevitably accompanied by a conformal alteration of $\Go$
($\Go' = e^{2 \lambda} \Go$).  This foretells that in \text{space-time-\!-time}
physics a nonvanishing electromagnetic potential field will produce real
effects even in regions where the electromagnetic field tensor vanishes,
a phenomenon already predicted by quantum mechanics \cite{13, 14}.
\vskip 15pt

\noindent {\bf VI.  Geodesic Equations}
\vskip 10pt

Let $p\: I \to {\eusm M}$ be a path in $\eusm M$, with parameter interval $I$,
and let the components of the velocity of $p$ in the adapted frame system
$\{ e_\mu, e_d \}$ be $\{ \pdot^\mu, \pdot^d \}$, so that
$\pdot = \pdot^\mu e_\mu (p) + \pdot^d e_d (p)$.  Then the acceleration
$\pbdotdot$ generated by the covariant differentiation $\bdhat$ is determined
by the connection forms of $\bdhat$, through use of Eqs\. (22), in the
following way:
$$
\align
\pbdotdot :\!@!@!@!@!@!@!@!
 &= (\pdot^\mu)\dot{}\, e_\mu (p)
    + \pdot^\kappa \bdhat e_\kappa (p) \pdot
    + \L(\pdot^d\R)\dot{}\, e_d (p)
    + \pdot^d \bdhat e_d (p) \pdot \\
\vspace{-7pt}
\tag30 \\
\vspace{-7pt}
 &= \pbdotdot^\mu e_\mu (p) + \pbdotdot^d e_d (p), \\
\intertext{where}
\pbdotdot^\mu
 &= (\pdot^\mu)\dot{}\, + \pdot^\kappa \omegahat_\kappa{}^\mu (p) \pdot
                        + \pdot^d \omegahat_d{}^\mu (p) \pdot
\tag31 \\
\intertext{and}
\pbdotdot^d
 &= \L(\pdot^d\R)\dot{}\, + \pdot^\kappa \omegahat_\kappa{}^d (p) \pdot
                                + \pdot^d \omegahat_d{}^d (p) \pdot.
\tag32
\endalign
$$
The condition that $p$ be an affinely parametrized geodesic path of $\bdhat$ is
that $\pbdotdot = 0$, which is equivalent to $\pbdotdot^\mu = \pbdotdot^d = 0$.
From Eqs\. (31), (32), (23), and (24), the fact that
$\omega^\lambda (p) \pdot = \pdot^\lambda$ and $\omega^d (p) \pdot = \pdot^d$,
and the skew-symmetry of $F_{\kappa \lambda}$ it follows that these geodesic
equations are equivalent, respectively, to
$$
\align
(\pdot^\mu)\dot{}\, + \pdot^\kappa \Gamma_\kappa{}^\mu{}_\lambda \pdot^\lambda
 &= \epshat \phi \pdot^d F^\mu{}_\lambda \pdot^\lambda
    - 2 \phi^{-1} \pdot^d \pdot^\mu
    + \epshat \pdot^d \pdot^d \phi^{-1} \phi^{. \mu}
\tag33 \\
\intertext{and}
\L(\pdot^d\R)\dot{}\, + \phi^{-1} \pdot^d \phi_{. \kappa} \pdot^\kappa
 &= \epshat \phi^{-1} \pdot^\kappa g_{\kappa \lambda} \pdot^\lambda,
\tag34
\endalign
$$
in which for brevity the compositions with $p$ of the various scalar fields
are implicit rather than express.

Utilizing Eqs\. (28) to break up $\Gamma_\kappa{}^\mu{}_\lambda$, and
remembering that $g_{\mu \nu} = e^{2 \zeta} \go _{\mu \nu}$ and
$g^{\nu \mu} = e^{-2 \zeta} \go^{\nu \mu}$, we find that Eqs\. (33) and
(34) are equivalent, respectively, to
$$
\align
(\pdot^\mu)\dot{}\, + \pdot^\kappa \Gammao_\kappa{}^\mu{}_\lambda \pdot^\lambda
 &= \epshat e^{-2 \zeta} \phi \pdot^d \Fo^\mu{}_\lambda \pdot^\lambda
    + 2 \L(\Ao_\kappa \pdot^\kappa - \phi^{-1} \pdot^d\R) \pdot^\mu\\
 &\qquad - \pdot^\kappa \go_{\kappa \lambda} \pdot^\lambda \Ao^\mu
         + \epshat e^{-2 \zeta} \pdot^d \pdot^d \phi^{-1} \phio^{. \mu}
\tag"($33{\bold '}$)" \\
\intertext{and}
\L(\pdot^d\R)\dot{}\, + \phi^{-1} \pdot^d \phi_{. \kappa} \pdot^\kappa
 &= \epshat e^{2 \zeta} \phi^{-1}
           \pdot^\kappa \go_{\kappa \lambda} \pdot^\lambda,
\tag"($34{\bold '}$)"
\endalign
$$
where $\Fo^\mu{}_\lambda := \go^{\kappa \mu} F_{\kappa \lambda}$ and
$\phio^{. \mu} := \phi_{. \lambda} \go^{\lambda \mu}$.  These equations
display explicitly all occurrences of $\zeta$ except those implied by
$\pdot^d = \omega^d (p) \pdot =
\L[\phi (\Ao_\kappa \omega^\kappa + d \zeta)\R](p) \pdot =
\phi (\Ao_\kappa \pdot^\kappa + \zetadot)$, where
$\zetadot := [\zeta (p)]\dot{\,}\, = d\zeta (p) \pdot$.  If we take this
decomposition of $\pdot^d$ partially into account, then we see that
Eq\. ($33'$) is equivalent to
$$
\align
\!\!\!\!\!\!\!\!\!\!\!\!\!\!\!\!\!\!
\L(e^{2 \zeta} \pdot^\mu\R)\dot{}\,
    + e^{2 \zeta} \pdot^\kappa \Gammao_\kappa{}^\mu{}_\lambda \pdot^\lambda
 &= \epshat \phi \pdot^d \Fo^\mu{}_\lambda \pdot^\lambda
     - e^{2 \zeta}
              \pdot^\kappa \go_{\kappa \lambda} \pdot^\lambda \Ao^\mu
     + \epshat \pdot^d \pdot^d \phi^{-1} \phio^{. \mu}.
\tag"($33{\bold {''}}$)" \\
\intertext{Noting further that $\phidot = \phi_{. \kappa} \pdot^\kappa$, we
find that Eq\. ($34'$) is equivalent to}
\L(\epshat \phi \pdot^d\R)\dot{}\,
 &= e^{2 \zeta} \pdot^\kappa \go_{\kappa \lambda} \pdot^\lambda.
\tag"($34{\bold {''}}$)"
\endalign
$$

As one knows, these geodesic equations entail that $\Ghat (p) \pdot \pdot$ is
constant.  This integral takes either of the equivalent forms
$$
\align
\pdot^\kappa g_{\kappa \lambda} \pdot^\lambda + \epshat \pdot^d \pdot^d
 &= \epsilon
\tag35 \\
\vspace{-3pt}
\intertext{and}
\vspace{-3pt}
e^{2 \zeta} \pdot^\kappa \go_{\kappa \lambda} \pdot^\lambda
   + \epshat \pdot^d \pdot^d
 &= \epsilon,
\tag"($35{\bold '}$)"
\endalign
$$
where $\epsilon := \sgn \L(\Ghat (p) \pdot \pdot\R) = 1$, $0$, or $-1$,
provided that the affine parametrization of $p$ is normal, that is,
that arclength is the parameter when $\Ghat (p) \pdot \pdot \neq 0$.
\vskip -7pt

$$
\align
\text{\noindent \!\!\bf VII.  }  &\text{\bf Momentum, Rest Mass, Electric
                                  Charge, Proper Time, and Equations of} \\
\vspace{-3pt}
                                 &\text{\bf Motion of a Test Particle in
                                  Space-Time-\!-Time}
\endalign
$$
\vskip -1pt

Thus far it has been convenient to leave unspecified both the dimensionality
$d$ of the manifold $\eusm M$ and the diagonal signature of the conformally
constrained metric $\Ghat$ carried by $\eusm M$.  Let us now restrict our
attention to the case in which $d = 5$ and $\Ghat$ is a
\text{space-time-\!-time} metric, with a view toward establishing a physical
interpretation of the \text{space-time-\!-time} geometry beyond that suggested
by comparison of it with its Weyl and Kaluza antecedents.  For this purpose it
is advantageous to have the signature of the space-time part of the metric be
$---+$; this causes the signature of $\Ghat$ to be $---++$ if $\epshat = 1$,
and to be $---+-$ if $\epshat = - 1$.

The procedure to be used here to effect a physical interpretation of the
geometry is a natural extension of the familiar space-time procedure.  One 
assumes that an elementary test particle's journey through life is described,
in whole or in part, by an affinely parametrized geodesic path $p$ in
\text{space-time-\!-time}.  One breaks the geodesic equation $\pbdotdot = 0$,
or some equivalent thereof, into its component equations in a perspicuously
appropriate frame system and compares these equations to the equations of
motion of a test particle in the special theory of relativity, or, more
closely, to the analogous equations of motion in the curved space-time of
general relativity theory.  Out of this comparison one identifies as far as
possible the various geometric parameters of the path $p$ with the classical
physical parameters of the particle.  In the same stroke one identifies terms
in the geodesic component equations as representing forces due to classical
physical fields, thus identifies the physical fields themselves with various of
the geometrical fields derived from the \text{space-time-\!-time} metric
$\Ghat$.  As this amounts to solving a puzzle in which no piece is seen to fit
until every piece is seen to do so, I shall dispense with many of the details
and go as quickly as possible to the conclusions.

To begin, let us define the {\bf \text{space-time-\!-time} momentum covector}
$P$ of the test particle to be the metric dual of its \text{space-time-\!-time}
velocity, that is, $P := \Ghat (p) \pdot$.  Because $\Ghat$ is
$\bdhat$-covariantly constant, $\Pbdot = \Ghat (p) \pbdotdot$, and therefore
the geodesic equation $\pbdotdot = 0$ is equivalent to $\Pbdot = 0$.  This
latter equation will provide the most immediate comparison to classical
equations of motion.  In the adapted coframe system
$\{ \omega^\mu, \omega^d \}$ the \text{space-time-\!-time} momentum $P$ has the
expansion $P = P_\kappa \omega^\kappa (p) + P_d \omega^d (p)$, where
$$
\align
P_\kappa
 &= \pdot^\mu g_{\mu \kappa} = e^{2 \zeta} \pdot^\mu \go_{\mu \kappa}
\tag36 \\
\vspace{-7pt}
\intertext{and}
\vspace{-7pt}
P_d
 &= \epshat \pdot^d.
\tag37 \\
\intertext{The covariant derivative of $P$ has the expansion
$\Pbdot = \Pbdot_\kappa \omega^\kappa (p) + \Pbdot_d \omega^d (p)$, where}
\Pbdot_\kappa
 &= (P_\kappa)\dot{}\, - P_\mu \Gamma_\kappa{}^\mu{}_\lambda \pdot^\lambda
                       - \phi P_d F_{\kappa \lambda} \pdot^\lambda
                       - \epshat P_d P_d \phi^{-1} \phi_{. \kappa}
\tag38 \\
\vspace{-7pt}
\intertext{and}
\vspace{-7pt}
\Pbdot_d
 &= (P_d)\dot{}\, + P_d \phi^{-1} \phi_{. \mu} \pdot^\mu
                  - \phi^{-1} P_\mu g^{\mu \nu} P_\nu
\tag39
\endalign
$$
(compositions of scalar fields with $p$ being suppressed in the notation), 
as follows from application of Eqs\. (22) and (23) to
$\Pbdot = (P_\kappa)\dot{}\, \omega^\kappa (p) +
P_\mu \bdhat \omega^\mu (p) \pdot + (P_d)\dot{}\, \omega^d (p) +
P_d \bdhat \omega^d (p) \pdot$.  Let
$$
\align
\mo :\!@!@!@!@!@!@!@!
 &= \L(\Go^{-1} (p) P P\R)^\frac12
  = \L(P_\mu \go^{\mu \nu} P_\nu\R)^\frac12 \\
\vspace{-7pt}
\tag40 \\
\vspace{-7pt}
 &= e^{2 \zeta} \L(\Go (p) \pdot \pdot\R)^\frac12
  = e^{2 \zeta} (\pdot^\mu \go_{\mu \nu} \pdot^\nu)^\frac12 \\ 
\vspace{-7pt}
\intertext{and}
\vspace{-7pt}
q :\!@!@!@!@!@!@!@!
 &= P \xi (p) = \phi P_d  \\
\vspace{-7pt}
\tag41 \\
\vspace{-7pt}
 &= \epshat \phi \pdot^d = \epshat \phi^2 A(p) \pdot
  = \epshat \phi^2 \L(\Ao_\mu \pdot^\mu + \zetadot\R).  
\endalign
$$
Then the equations $\Pbdot_\kappa = 0$ and $\Pbdot_d = 0$, equivalent jointly
to $\Pbdot = 0$, are equivalent respectively to
$$
\align
(P_\kappa)\dot{}\,
 &= P_\mu \Gammao_\kappa{}^\mu{}_\lambda \pdot^\lambda
    + q F_{\kappa \lambda} \pdot^\lambda
    - e^{-2 \zeta} \mo^2 \Ao_\kappa
    + \epshat (q / \phi)^2 \phi^{-1} \phi_{. \kappa} \\
\vspace{-5pt}
\tag42 \\
\vspace{-5pt}
 &= e^{-2 \zeta}
      P_\mu \Gammao_\kappa{}^\mu{}_\lambda P_\nu \go^{\nu \lambda}
      + e^{-2 \zeta} q F_{\kappa \lambda} P_\nu \go^{\nu \lambda}
      - e^{-2 \zeta} \mo^2 \Ao_\kappa
      + \epshat (q / \phi)^2 \phi^{-1} \phi_{. \kappa} \\
\vspace{-7pt}
\intertext{and}
\vspace{-7pt}
\qdot
 &= e^{-2 \zeta} \mo^2.
\tag43
\endalign
$$
These equations have, if the affine parametrization of $p$ is normal,
the integral $\Ghat^{-1} (p) P P = \epsilon$.  This is, of course,
the same as $\Ghat (p) \pdot \pdot = \epsilon$, and therefore the same as
Eq\. ($35'$), which is equivalent in terms of $\mo$ and $q$ to
$$
e^{-2 \zeta} \mo^2 + \epshat (q / \phi)^2 = \epsilon.
\tag44
$$
Substitution of this integral into Eq\. (43) yields
$$
\align
\qdot
 &= \epsilon - \epshat (q / \phi)^2.
\tag45 \\
\vspace{-7pt}
\intertext{Equations (44), (41), and (43) imply that}
\vspace{-7pt}
\L(\mo^2\R)\dot{}\,
 &= 2 \L[-\mo^2 \Ao_\kappa
            + \epshat e^{2 \zeta}
              (q / \phi)^2 \phi^{-1} \phi_{. \kappa}\R] \pdot^\kappa \\
\vspace{-7pt}
\tag46 \\
\vspace{-7pt}
 &= 2 \L[-e^{-2 \zeta} \mo^2 \Ao_\kappa
            + \epshat (q / \phi)^2 \phi^{-1} \phi_{. \kappa}\R]
                                              P_\lambda \go^{\lambda \kappa}.
\endalign
$$

The scalar $\Go (p) \pdot \pdot$, otherwise identifiable as
$\pdot^\mu \go_{\mu \nu} \pdot^\nu$ and as $e^{-4 \zeta} \mo^2$, may be
positive, zero, or negative on different geodesics and, generally, on different
portions of the same geodesic.  It is the square length of the
``space-time part'' $\pdot^\mu e _\mu (p)$ of the velocity $\pdot$, as measured
by the degenerate metric $\Go$, whose space-time part has diagonal signature
$---+$.  Wherever on $p$ this scalar is positive, that is, wherever the
space-time part of $\pdot$ is timelike, we can introduce a real parameter
$\tauo$ such that
$$
\align
\tauo :\!@!@!@!@!@!@!@!
 &= \int \L(\Go (p) \pdot \pdot\R)^\frac12 \, d \tauhat
  = \int (\pdot^\mu \go_{\mu \nu} \pdot^\nu)^\frac12 \, d \tauhat \\
\vspace{-7pt}
\tag47 \\
\vspace{-7pt}
 &= \int e^{-2 \zeta} \mo \, d \tauhat = \int \mo^{-1} \, dq 
\endalign
$$
($\tauhat$ denoting the \text{space-time-\!-time} affine parameter of $p$), and
with it define space-time velocity components $u^\lambda$ by
$u^\lambda \! := dp^\lambda / d\tauo \! := \pdot^\lambda / (\tauo )\dot{}\,$.
Equations (42), (43), and (46) then are equivalent, wherever $\mo^2 > 0$, to
$$
\align
\frac{dP_\kappa}{d\tauo}
 &= P_\mu \Gammao_\kappa{}^\mu{}_\lambda u^\lambda
    + q F_{\kappa \lambda} u^\lambda
    - \mo \Ao_\kappa
    + \epshat e^{2 \zeta} \mo^{-1} (q / \phi)^2 \phi^{-1} \phi_{. \kappa},
\tag"($42 {\bold '}$)" \\
\frac{dq}{d\tauo}
 &= \mo,
\tag"($43 {\bold '}$)"
\endalign
$$
and
$$
\frac{d(\mo^2)}{d\tauo} = 2 \L[-\mo^2 \Ao_\kappa
                                  + \epshat e^{2 \zeta} (q / \phi)^2 
                                    \phi^{-1} \phi_{. \kappa}\R] u^\kappa.
\tag"($46 {\bold '}$)"
$$

Upon comparing these equations with the classical relativistic equations of
motion for an electrically charged particle, and remembering the various
definitions that have gone into them, one arrives at the following 
identifications and conclusions:
\item{1.}
The scalar parameter $\tauo$ is a (space-time) proper time parameter of the
particle.
\item{2.}
The $u^\lambda$ are the components of the space-time proper velocity vector
of the particle.
\item{3.}
The $P_\kappa$ are the components of the space-time momentum covector of the
particle.
\item{4.}
The scalar parameter $\mo$ is the rest mass of the particle.
\item{5.}
The scalar parameter $q$ is the electric charge of the particle.
\item{6.}
The $F_{\kappa \lambda}$ are the components of the space-time electromagnetic
field tensor.
\item{7.}
The $\Ao_\kappa$ are the components of a space-time covector potential field
for the electromagnetic field.
\item{8.}
The apparent forces to which the particle is subject, in that they contribute,
according to Eq\. ($42'$), additively to the space-time momentum rates
$dP_\kappa / d\tauo$, consist of
\itemitem{a.}
the gravitational and other forces attributable to space-time geometry that are
included in the term $P_\mu \Gammao_\kappa{}^\mu{}_\lambda u^\lambda$, familiar
from general relativity theory;
\itemitem{b.}
the Lorentz force of the electromagnetic field, expressed by the term
$q F_{\kappa \lambda} u^\lambda$;
\itemitem{c.}
a rest-mass proportional force in the direction of the electromagnetic
potential, expressed by the term $-\mo \Ao_\kappa$; and
\itemitem{d.}
a force proportional to the square of the electric charge, inversely 
proportional to the rest mass, and in the direction of the gradient of the
scalar field $\phi$, expressed by the term
$\epshat e^{2 \zeta} \mo^{-1} (q / \phi)^2 \phi^{-1} \phi_{. \kappa}$.
\item{9.}
Neither the electric charge $q$ nor the rest mass $\mo$ can be expected in
general to remain constant, as they will evolve in accordance with
Eqs\. $(43')$ and $(46')$ while maintaining a kind of joint conservation,
described by Eq\. (44).
\vskip 5pt

To go one step further, let $\Po^\mu := P_\nu \go^{\nu \mu}$.  Then
$\Po^\mu = e^{2 \zeta} \pdot^\mu$, and $\Po^\mu = \mo u^\mu$ wherever
$\mo^2 > 0$, in consequence of which we may identify the $\Po^\mu$ as the
components of the space-time momentum vector of the particle.  Consistent with
this identification is the observation that
$\mo^2 = \Po^\mu \go_{\mu \nu} \Po^\nu$.  In terms of $\Po^\mu$, $q$, and
$\mo$, the geodesic equation $(33'')$ reads
$$
\align
\!\!\!\!\!\!
\L(\Po^\mu\R)\dot{}\,
 + e^{-2 \zeta} \Po^\kappa \Gammao_\kappa{}^\mu{}_\lambda \Po^\lambda
 &= e^{-2 \zeta} q \Fo^\mu{}_\lambda \Po^\lambda
     - e^{-2 \zeta} \mo^2 \Ao^\mu
     + \epshat (q / \phi)^2 \phi^{-1} \phio^{. \mu}.
\tag48 \\
\vspace{-5pt}
\intertext{And this is equivalent, wherever $\mo^2 > 0$, to}
\vspace{-3pt}
\!\!\!\!\!\!
\frac{d(\mo u^\mu)}{d\tauo}
  + \mo u^\kappa \Gammao_\kappa{}^\mu{}_\lambda u^\lambda
 &= q \Fo^\mu {}_\lambda u^\lambda
     - \mo \Ao^\mu
     + \epshat e^{2 \zeta} \mo^{-1} (q / \phi)^2 \phi^{-1} \phio^{. \mu},
\tag"($48 {\bold '}$)"
\endalign
$$
an equation which helps to cement the identifications and conclusions outlined
above.

As one knows, the nonnull geodesic paths of $\bdhat$ are the paths that 
make stationary the arclength integral
$\int_{\tauhat_1}^{\tauhat_2} |\pdot|\, d\tauhat$, in which
$|\pdot| := |\Ghat (p) \pdot \pdot|^\frac12$.  The canonical momentum
covector $M$ whose components appear in the Euler equations for this
variational problem can be expressed by
\vskip -3pt
$$
\aligned
M :\!@!@!@!@!@!@!@!
 &= (\partial |\pdot| / \partial \pdot^\kappa) \omega^\kappa (p)
    + (\partial |\pdot| / \partial \pdot^d) \omega^d (p) \\
 &= \sgn \L(\Ghat (p) \pdot \pdot\R) |\pdot|^{-1} \Ghat (p) \pdot \\ 
 &= \epsilon |\pdot|^{-1} P
    = \epsilon |\pdot|^{-1} [P_\kappa \omega^\kappa (p) + P_d \omega^d (p)].
\endaligned
\tag49
$$
From this it follows that $P_\kappa = \epsilon |\pdot| M_\kappa =
(1/2) (\partial L / \partial \pdot^\kappa)$ and
$P_d = \epsilon |\pdot| M_d = (1/2) (\partial L / \partial \pdot^d)$, where
$L := \epsilon |\pdot|^{2} = \Ghat (p) \pdot \pdot$, and that the equations
of motion (42) and (43) (which, being equivalent to $\pbdotdot = 0$, hold only
for affine parametrizations of $p$) can be derived from an action principle
with $L$ as the Lagrangian \cite{15}.  In terms of $\mo$ and $q$ this
Lagrangian can be formulated thus:
\vskip -2pt
$$
\aligned
L &= e^{-2 \zeta} \mo^2 + \epshat (q / \phi)^2 \\ 
  &= \mo (v^\mu \go_{\mu \nu} v^\nu)^\frac12 + q \Ao_\mu v^\mu + q \zetadot;
\endaligned
\tag50
$$
here $v^\mu := \pdot^\mu$ and Eqs\. (40) and (41) have been invoked.
But for the extra term $q \zetadot$, which refers to progression along the
secondary time dimension and therefore has no space-time analog, this
\text{space-time-\!-time} Lagrangian would duplicate in appearance a standard
space-time Lagrangian for the equations of motion of a charged particle in the
special theory of relativity \cite{16} and, by simple extension, in the general
theory as well.  In assessing this correspondence, however, one should bear in
mind that in \text{space-time-\!-time} $\mo$ and $q$ are geometric parameters
of the geodesic, not, as in space-time theories, mere handcrafted constants of
no geometrical significance.

It is clear that the Lagrangian $L$, the geodesic equation $\pbdotdot = 0$,
and its equivalent $\Pbdot = 0$ are all gauge invariant, inasmuch as gauge
transformations are just coordinate transformations (of the type
$\[[ x^{\mu'}, \zeta \]] \to \[[ x^{\mu'}, \zeta - \lambda \]]$), which do
not affect $\Ghat$, $\pdot$, $\pbdotdot$, $P$, or $\Pbdot$.  What is not so
apparent is that each component equation of motion is individually gauge
invariant.  This comes about because $\{ \omega^\mu, \omega^d \}$, consequently
$\{ e_\mu, e_d \}$, and therefore $\Pbdot_\mu$, $\Pbdot_d$, $\pbdotdot^\mu$,
and $\pbdotdot^d$, stay fixed when the gauge changes (as, likewise,
do $P_\mu$, $P_d$, $\pdot^\mu$, and $\pdot^d$).  Thus, every one of
Eqs\. (42), (43), ($42'$), ($43'$), (48), and ($48'$) is individually gauge
invariant (up to simple algebraic equivalence).  Also gauge invariant is the
electric charge $q$, as follows from the fact that
$q = \epshat \phi^2 A(p) \pdot$, no part of which is altered by a change of
gauge.  Not gauge invariant, however, are the rest mass $\mo$ and the proper
time $\tauo$, which when $\zeta \to \zeta - \lambda$ behave so:
$\mo \to e^{-\lambda} \mo$ and
$d\tauo / d\tauhat \to e^{\lambda} (d\tauo / d\tauhat)$.  Nor are the
components $\Po^\mu$ ($= \mo u^\mu$) of the space-time momentum vector gauge
invariant, for $\Po^\mu \to e^{-2 \lambda} \Po^\mu$.  The product
$\mo (d\tauo / d\tauhat)$, however, is gauge invariant, as is
$\sgn (\mo^2)$.  The lack of invariance for $\mo$, $\tauo$, and $\Po^\mu$ of
course reflects the fact that in the new gauge it is $e^{2 \lambda} \Go$
instead of $\Go$ that is considered to be the metric of space-time.

Test particles obeying the equations of motion here detailed exhibit a
complexity of behavior far beyond that of test particles in Einstein's
space-time theory or in its extensions by Weyl, Kaluza, Klein, and others.
This is owed in large measure to the unprecedented manner in which the
electric charge $q$ evolves and the equally unprecedented nature of the
coupling of momentum rates to the gradient of $\phi$.  These have among their
effects that a test particle can appear (seemingly out of nowhere) at a
space-time event ${\eusm E}_1$ with $q = -\phi ({\eusm E}_1)$ and vanish at a
later event ${\eusm E}_2$ with $q = \phi ({\eusm E}_2)$, and that {\it at}
${\eusm E}_1$ and {\it at} ${\eusm E}_2$ the $\phi$-gradient force will,
because of the growth of the coupling factor $e^{2 \zeta}$ in Eq\. ($42'$),
{\it infinitely} dominate the other forces and thereby draw the particle
irresistibly into the depths of one of the potential wells of $\epshat \phi$.
These potential wells thus are the most probable locations for the occurrence
of such ``creation'' and ``annihilation'' events.  The thought that such
behavior might be used to model orbital transitions (``quantum jumps'') of
electrons in atoms cannot be suppressed.

Because of its complexity I shall not here attempt further to describe
\text{space-time-\!-time} test particle behavior.  Instead, I shall, in the
next section, discuss subtleties in the concepts of mass and of charge that
flow from these equations of motion, subtleties involving distinctions often
unmade or neglected --- to the detriment of science, for to fail to distinguish
is to fail to know.
\vskip 15pt

\noindent {\bf VIII.  The Inertial-Passive Equivalence and the Passive-Active
                      Distinction}
\vskip 10pt

In Newton's theory of gravity the assumption that a test particle's inertial
mass $m_i$ and its passive gravitational mass $m_p$ are equal (and constant)
reduces the equation of motion
$(m_i \dot {\bold r})\dot{}\, + (m_p M / r^2)(\bold r / r) = \bold 0$ to the
equation $\ddot {\bold r} + (M / r^2)(\bold r / r) = \bold 0$, in which neither
of those masses appears.  Einstein's theory of gravity incorporates that same
equivalence by admitting only space-time geodesics as worldlines of test
particles.  It thereby adopts as its equation of motion a generalization of
the {\it reduced} Newtonian equation, thus avoids even introducing $m_i$ and
$m_p$ as concepts of significance for gravity.  Because test particles in
\text{space-time-\!-time} must deal with the electromagnetic field alongside
the gravitational field, this theory cannot exclude those concepts.  It
introduces them effortlessly, however, and in such a way as to maintain the
numerical equivalence of $m_i$ and $m_p$ and to make them ignorable in the
absence of nongravitational fields.  Specifically, the same mass parameter
$\mo$ that appears in the first term of Eq\. ($48'$) in the role of inertial
(rest) mass $m_i$ appears also in the second term in the role of passive
gravitational mass $m_p$; thus in \text{space-time-\!-time}
$m_i := \mo =: m_p$.  And when the nongravitational fields
$\phi_{. \kappa}$, $\Ao_\kappa$, and $F_{\kappa \lambda}$ are all zero, then
the horizontal subspaces are (space-time) hypersurface-forming, and $\mo$ has
to be constant to satisfy Eq\. $(46')$, whereupon Eq\. ($48'$) reduces to
$du^\mu / d\tauo + u^\kappa \Gammao_\kappa{}^\mu{}_\lambda u^\lambda = 0$,
which implies that the particle's space-time trajectory is geodesic, just as
in Einstein's theory.

The constant $M$ in the Newtonian equations of motion tells the strength
of the gravitational field acting on the test particle; it is properly
called the {\it active} gravitational mass of the particle considered to be
producing that field, which of course is not the test particle.  Newtonian
theory treats every particle as both a test particle with $m_i = m_p$ and a
field-generating particle with an active gravitational mass $m_a$.  Although
$m_a$ and $m_p$ refer to entirely different concepts, Newton's law of action
and reaction, {\it applied instantaneously at a distance}, allows the
inference that $m_a = m_p$.

In Einstein's theory the analog of $M$ is the Schwarzschild mass parameter
$M_S$, which also is properly called the active gravitational mass of the
``particle'' whose gravitational field the Schwarzschild metric represents.
Although as noted that theory has no concept of passive gravitational mass,
one can insert $m_p$ and its equal $m_i$ into the radial equation of geodesic
motion for a Schwarzschild metric at the expected places to obtain an
equivalent equation generalizing the unreduced Newtonian equation, with $M_S$
in place of $M$.  This done, however, one yet finds it impossible to establish
by the Newtonian argument any equivalence between active and passive mass
parameters.  Even if the logically chimerical notion of a test particle with
an active gravitational mass $m_a$ as well as a passive gravitational mass
$m_p$ be entertained, the Newtonian argument that $m_a = m_p$ founders on the
lack of any ``instantaneous gravitational action and reaction at a distance''
in Einstein's theory.

In \text{space-time-\!-time} theory the situation is the same:  there is no
concept of an active gravitational mass of a test particle; an analog of the
Newtonian $M$ and the Schwarzschild\-ean $M_S$ can exist in a particular
\text{space-time-\!-time}, but it is a parameter of the gravitational field of
that \text{space-time-\!-time}, not of any test particle that the field acts
upon; if particles with both active and passive gravitational masses be
imagined, then the finiteness of the speed of propagation of gravitational
effects precludes establishment of any relationship between the two masses.
But in this theory a further, similar discrimination is unavoidable.  The
electric charge parameter $q$ of a \text{space-time-\!-time} test particle
measures, in its initial appearance in Eq\. ($48'$), the response of the
particle to the electromagnetic field $F_{\kappa \lambda}$.  Thus it plays
there the role of a {\it passive} electric charge, just as $\mo$ takes the role
of a passive gravitational mass in its second appearance in that equation.  If
$F_{\kappa \lambda}$ should have a form like that of a Coulomb field of
strength $Q$, then $Q$ would properly be called the {\it active} electric
charge of the particle considered to be generating that field, but that
particle could not strictly be treated as a test particle at all, still less as
a test particle with passive electric charge $Q$.  Between these concepts of
active and of passive electric charge, just as between the concepts of active
and of passive gravitational mass, lies a broad gulf, across which no bridge is
apparent.  Essentially the same gulf is present already in Maxwell--Lorentz
electrodynamics.  Attempts to bridge it there, by supposing test particles to
have active (or at least semi-active) charge as well as passive charge, have
produced among other oddities an equation of motion with an $\bdrdotdotdot$
term that lets in self-accelerated ``runaway'' solutions.  The
\text{space-time-\!-time} equation of motion ($48'$) has no comparable term and
no such solution.  \text{Space-time-\!-time} theory seems to require no
bridge across the active-passive electric charge gulf, or for that matter
across the active-passive gravitational mass gulf.  It is conceivable, however,
that some such connections lie hidden in the theory, to be exposed by future 
investigation \cite{17}.

In its third appearance in Eq\. ($48'$) $\mo$ helps to measure the response of
the test particle to the field $\Ao_\mu$, and in the last term $q$ and $\mo$
combine to help determine the particle's response to the field $\phi_{. \mu}$.
The apparent forces involved are peculiar to \text{space-time-\!-time}, so
there are no names like ``passive gravitational mass'' and ``passive electric
charge'' ready at hand to signify the roles played here by $\mo$ and
$q^2 / \mo$.  This is perhaps fortunate, for such names tend to mislead by
putting attention on the apparent forces themselves, rather than on the
underlying geometry they spring from.  It is this geometry that is presumed to
model reality; the apparent forces and the test particles following geodesics
are just convenient fictions to help us connect the geometry to our
perceptions.
\vskip 15pt

\noindent {\bf IX.  Curvature}
\vskip 10pt

A full physical interpretation of the geometry of \text{space-time-\!-time}
must rest ultimately not only on delineation of the mechanics of test
particles, but also on establishment of field equations for the evolution and
interactions of $\phi$, $\Ao$, $F$, and $\Go$, analysis of the field dynamics
those equations imply, and arrival at an understanding of the physical import
of the unfamiliar scalar field $\phi$.  In preparation for a subsequent paper
deriving such field equations I shall exhibit here and in the Appendix both 
concise and not so concise forms of the curvature tensor field $\Thetahat$ of
the conformally constrained metric $\Ghat$, its contracted curvature tensor
field $\Phihat$, its curvature scalar field $\Psihat$, and its Einstein tensor
field $\Ehat$.  The adapted frame system $\{e_\mu, e_d\}$ and its dual
$\{\omega^\mu, \omega^d\}$ are best suited to this purpose.  As earlier, no
restriction is placed on the dimensionality of $\eusm M$ or the signature of
$\Ghat$.

If we adopt the convention that $K$, $L$, $M$, $N$, etc\. range from 1 to $d$
(retaining for $\kappa$, $\lambda$, $\mu$, $\nu$, etc\. the range 1 to
$d - 1$), then we have that
$\Thetahat = \omega^K \lotimes \Thetahat_K{}^M \lotimes e_M$, with the
curvature 2-forms $\Thetahat_K{}^M$ computed from the connection 1-forms of
Eqs\. (23) by means of the structural equation $\Thetahat_K{}^M =
2(d_\wedge \omegahat_{K}{}^M - \omegahat_{K}{}^P \lwedge \omegahat_{P}{}^M)$.
Upon performing the computations one finds that \cite{11}
$$
\align
\Thetahat_\kappa{}^\mu
 &= \Theta_\kappa{}^\mu
{\topaligned
  &- \L[\epshat \phi^2 F_\kappa{}^{(\mu} F_{\lambda) \nu}
   + \epshat 2 \phi^{-2} g_{\kappa \lambda} g^\mu{}_\nu\R]
                                        \omega^\nu \lwedge \omega^\lambda \\
  &+ \L[\epshat \L(\phi F_\kappa{}^\mu{}_{; \nu}
            + 2F_{(\kappa}{}^\mu \phi_{. \nu )}
            + 2F_\kappa{}^{(\mu} \phi_{. \nu )}\R)
            + 4 \phi^{-2} \phi_{. [\kappa} g^{\mu]}{}_\nu\R]
                                           \omega^\nu \lwedge \omega^d,
\endtopaligned} \\
\Thetahat_\kappa{}^d
 &= \L[\phi F_{\kappa \lambda ; \nu}
     - 2 \phi_{. (\kappa} F_{\lambda) \nu}
     + \epshat 2 \phi^{-2} g_{\kappa \lambda} \phi_{. \nu}\R]
                                     \omega^\nu \lwedge \omega^\lambda \\
\vspace{-7pt}
\tag51 \\
\vspace{-7pt}
 &\qquad - \l[2 \phi^{-1} \phi_{. \kappa ; \lambda}
   + \epshat (1/2) \phi^2 F_\kappa{}^\pi F_{\pi \lambda}
   + \epshat 2 \phi^{-2} g_{\kappa \lambda}\r]
                                        \omega^d \lwedge \omega^\lambda, \\
\Thetahat_d{}^\mu
 &= - \epshat g^{\mu \kappa} \Thetahat_\kappa \\
\vspace{-14pt}
\intertext{and}
\vspace{-14pt}
\Thetahat_d{}^d
 &= 0.
\endalign
$$
In the first of these equations
$$
\align
\Theta_\kappa{}^\mu :\!@!@!@!@!@!@!@!
 &= [2 (\Gamma_\kappa{}^\mu{}_{\lambda , \nu}
         + \Gamma_\kappa{}^\pi{}_\lambda \Gamma_\pi{}^\mu{}_\nu 
         + \Gamma_\kappa{}^\mu{}_\pi C_\lambda{}^\pi{}_\nu) \\
 & \qquad - (g_\kappa{}^\mu F_{\lambda \nu}
                 - F_{\kappa \lambda} g^\mu{}_\nu
                 - g_{\kappa \lambda} F^\mu{}_\nu)] \,
                                       \omega^\nu \lwedge \omega^\lambda.
\tag52
\endalign
$$
The other abbreviations introduced in them are
$$
\align
\phi_{. \kappa ; \lambda}
 &:= \phi_{. \kappa, \lambda}
      - \phi_{. \pi} \Gamma_\kappa{}^\pi{}_\lambda, \\
F_{\kappa \lambda ; \nu}
 &:= F_{\kappa \lambda , \nu}
     - F_{\kappa \pi} \Gamma_\lambda{}^\pi{}_\nu
     - F_{\pi \lambda} \Gamma_\kappa{}^\pi{}_\nu,
\tag53 \\
\vspace{-14pt}
\intertext{and}
\vspace{-14pt}
F_\kappa{}^\mu{}_{; \nu}
 &:= F_\kappa{}^\mu{}_{, \nu}
     + F_\kappa{}^\pi \Gamma_\pi{}^\mu{}_\nu
     - F_\pi{}^\mu \Gamma_\kappa{}^\pi{}_\nu.
\endalign
$$

The ``${}_;$'' operation of Eqs\. (53) harks back to the covariant
differentiation $\bold d$ defined in Sec\. V, for which $\bold d e_\kappa =
\omega_\kappa{}^\mu \otimes e_\mu$, $\bold d e_d = 0$, $\bold d \omega^\mu =
- \omega_\kappa{}^\mu \otimes \omega^\kappa$, and $\bold d \omega^d = 0$,
with $\omega_\kappa{}^\mu = \Gamma_\kappa{}^\mu{}_\lambda \omega^\lambda$.
Thus, $\phi$ being independent of $\zeta$, $d \phi =
\phi_{, \kappa}\omega^\kappa = \phi_{. \kappa} \omega^\kappa$
and $d \phi_{. \kappa} = \phi_{. \kappa , \lambda} \omega^\lambda$, so
$\bold d (d \phi) =
d \phi_{. \kappa} \otimes \omega^\kappa + \phi_{. \pi} \bold d \omega^\pi =
(d \phi_{. \kappa} - \phi_{. \pi} \omega_\kappa{}^\pi) \otimes \omega^\kappa =
\phi_{. \kappa ; \lambda} \omega^\lambda \otimes \omega^\kappa$.  A similar
calculation finds that $\bold d F = F_{\kappa \lambda ; \nu}
\omega^\nu \otimes \omega^\lambda \otimes \omega^\kappa$.  On the other hand
$F_\kappa{}^\mu$ is not independent of $\zeta$, so $\bold d (FG^{-1} ) =
\bold d (F_\kappa{}^\mu e_\mu \otimes \omega^\kappa) -
(F_\kappa{}^\mu{}_{; \nu} \omega^\nu + F_\kappa{}^\mu{}_{;d} \omega^d) \otimes
e_\mu \otimes \omega^\kappa$, where, because all of $\bold d$'s connection
coefficients other than the $\Gamma_\kappa{}^\mu{}_\lambda$ vanish,
$F_\kappa{}^\mu{}_{;d} := F_\kappa{}^\mu{}_{,d}$.  It is easy to see that
$g_{\mu \nu ; \lambda} = g^{\mu \nu}{}_{; \lambda} = 0$ and that
$F_\kappa{}^\mu{}_{; \nu} = g^{\mu \lambda} F_{\kappa \lambda ; \nu}$.
Should the need arise, Eqs\. (53) can be refigured by use of the equivalences
$\phi_{. \kappa , \lambda} = \phi_{. \kappa . \lambda} =:
\phi_{. \kappa \lambda}$, $F_{\kappa \lambda , \nu} =
F_{\kappa \lambda . \nu}$, and $F_\kappa{}^\mu{}_{, \nu} =
F_\kappa{}^\mu{}_{. \nu} + 2 F_\kappa{}^\mu \Ao_\nu$.
The notation $\Theta_\kappa{}^\mu$ notwithstanding, the 2-forms of Eq\. (52)
are not curvature forms of $\bold d$, nor would they be if only the terms
involving $\Gamma$ were present.

By taking account of the skew-symmetries involved, one can extract from
Eqs\. (51) the curvature components $\Thetahat_K{}^M\!{}_{LN}$ that appear in
$\Thetahat_K{}^M = \Thetahat_K{}^M\!{}_{LN} \omega^N \lwedge \omega^L =
\Thetahat_K{}^M\!{}_{LN} \omega^N \lotimes \omega^L$.  This is done in the
Appendix.

The contracted curvature tensor field $\Phihat$ can be computed directly
from Eqs\. (51) by use of $\Phihat := \omega^R \Thetahat (\bold \cdot) e_R =
\omega^K \lotimes \omega^R( \Thetahat_K{}^M \lotimes e_M) e_R =
\omega^K \lotimes \Thetahat_K{}^R e_R$, or, by referring to Eqs\. (A.1) in
the Appendix, from
$\Phihat = \omega^K \lotimes \Thetahat_K{}^R\!{}_{LR} \omega^L$.  The result
is that $\Phihat = \omega^K \lotimes \Phihat_{KL} \omega ^L$, where
$$
\align
\Phihat_{\kappa \lambda}
 &= \Phi_{\kappa \lambda}
     - \phi^{-1} \phi_{. \kappa ; \lambda}
     + \epshat (1/2) \phi^2 F_\kappa{}^\rho F_{\rho \lambda}
     - \epshat (d - 1) \phi^{-2} g_{\kappa \lambda}, \\
\Phihat_{\kappa d}
 &= \epshat (1/2) (\phi F_\kappa{}^\rho{}_{; \rho}
                    + 3F_\kappa{}^\rho \phi_{. \rho})
                    + (d - 2) \phi^{-2} \phi_{. \kappa}, \\
\Phihat_{d \lambda}
 &= \Phihat_{\lambda d},
\tag54 \\
\vspace{-14pt}
\intertext{and}
\vspace{-14pt}
\Phihat_{dd}
 &= - \epshat \phi^{-1} \phi^{. \rho}{}_{; \rho}
    - (1/4) \phi^2 F^\rho{}_\pi F^\pi{}_\rho
    - (d - 1) \phi^{-2}.
\endalign
$$
In the first of these equations
$$
\align
\Phi_{\kappa \lambda} :\!@!@!@!@!@!@!@!
 &= \Theta_\kappa{}^\rho e_\rho e_\lambda
  = \Theta_\kappa{}^\rho{}_{\lambda \rho} \\ 
\vspace{-7pt}
\tag55 \\
\vspace{-7pt}
 &= 2 \l(\Gamma_\kappa{}^\rho{}_{[\lambda , \rho]}
          + \Gamma_\kappa{}^\pi{}_{[\lambda} \Gamma_\pi{}^\rho{}_{\rho]}
          + \Gamma_\kappa{}^\rho{}_\pi C_\rho{}^\pi{}_\lambda\r)
    + (1/2)(d - 1) F_{\kappa \lambda}, \\
\vspace{-7pt}
\intertext{and in the last}
\vspace{-7pt}
\phi^{. \rho}{}_{; \rho} :\!@!@!@!@!@!@!@!
 &= \phi^{. \rho}{}_{, \rho} + \phi^{. \pi} \Gamma_\pi{}^\rho{}_\rho
  = \phi_{. \kappa ; \rho} g^{\kappa \rho}.
\tag56
\endalign
$$

Because $\Ghat^{-1} =
e_\mu \otimes g^{\mu \nu} e_\nu + \epshat e_d \otimes e_d$, we have that
$\Ghat^{-1} \Phihat = \omega^\kappa \otimes \Phihat_\kappa{}^\nu e_\nu
+ \omega^d \otimes \epshat \Phihat_{dd} e_d$, with
$\Phihat_\kappa{}^\nu := \Phihat_{\kappa \lambda} g^{\lambda \nu}$, hence
that $\Psihat := \omega^P \L(\Ghat^{-1} \Phihat\R) e_P =
\Phihat_\pi{}^\pi + \epshat \Phihat_{dd}$.  Applying this to Eqs\. (54)
one finds that
$$
\Psihat = \Psi - 2 \phi^{-1} \phi^{. \rho}{}_{; \rho}
               + \epshat (1/4) \phi^2 F_\pi{}^\rho F_\rho{}^\pi
               - \epshat (d - 1)d \,\phi^{-2},
\tag57 
$$
where
$$
             \Psi := \Phi_\pi{}^\pi \quad \text{and} \quad
\Phi_\kappa{}^\nu := \Phi_{\kappa \lambda} g^{\lambda \nu}.
\tag58
$$
It then follows from $\Ehat := \Phihat - (1/2) \Psihat \Ghat$ that
$\Ehat = \omega^K \lotimes \Ehat_{KL} \omega^L$, where
$$
\align
\Ehat_{\kappa \lambda}
 &= E_{\kappa \lambda}
{\topaligned
  &- \phi^{-1} (\phi_{. \kappa ; \lambda}
                 - \phi^{. \rho}{}_{; \rho} g_{\kappa \lambda}) \\
  &+ \epshat (1/2) \phi^2
      [F_\kappa{}^\rho F_{\rho \lambda}
       - (1/4) F_\pi {}^\rho F_\rho {}^\pi g_{\kappa \lambda}] \\
  &+ \epshat (1/2)(d-2)(d-1) \phi^{-2} g_{\kappa \lambda},
\endtopaligned} \\
\Ehat_{\kappa d}
 &= \epshat (1/2) (\phi F_\kappa{}^\rho{}_{; \rho}
                    + 3 F_\kappa{}^\rho \phi_{. \rho})
     + (d-2) \phi^{-2} \phi_{. \kappa},
\tag59 \\
\Ehat_{d \lambda}
 &= \Ehat_{\lambda d}, \\
\vspace{-14pt}
\intertext{and}
\vspace{-14pt}
\Ehat_{dd}
 &= - \epshat (1/2) \Psi - (3/8) \phi^2 F_\pi{}^\rho F_\rho{}^\pi
                         + (1/2)(d-2)(d-1) \phi^{-2}.
\endalign
$$
The abbreviation
$$
E_{\kappa \lambda} := \Phi_{\kappa \lambda} - (1/2) \Psi g_{\kappa \lambda}
\tag60
$$
is used in the first of Eqs\. (59).

Hidden within these relatively concise expressions of $\Thetahat$, $\Phihat$,
$\Psihat$, and $\Ehat$ is a wealth of ``interactions'' among the fields 
$\phi$, $\Ao_\mu$, $F_{\mu \nu}$, and $\go _{\mu \nu}$.  To bring them to
visibility we shall have to ``detelescope'' the expressions with the aid of the
expansions set out in Eqs\. (25)--(29).  The procedure is straightforward, but
the product will occupy a considerable space.  To reduce congestion the
detelescoped  expressions for $\Thetahat$ and $\Phihat$ will be displayed in
the Appendix, leaving here only those for $\Psihat$ and $\Ehat$.  In both
places will appear additional abbreviations which can be described in the
following way:  The practice of inserting a ``$\; \degree \;$'' to indicate
raising of an index with the $\go^{\mu \nu}$ rather than the $g^{\mu \nu}$
(as in $\phio^{. \mu} := \phi_{. \lambda} \go^{\lambda \mu}$) is continued,
and is sharpened by the stipulations that if a ``$\; \degree \;$'' is already
present, then {\it only} the $\go^{\mu \nu}$ can raise an index, and that
$\Ao^\mu := \Ao_\lambda \go^{\lambda \mu}$, {\it not}
$A_\lambda \go^{\lambda \mu}$).  Further, it is understood that the
``$\; \degree \;$'' travels with a raised index involved in a symmetrization 
or an antisymmetrization; thus $\Fo_\kappa{}^{( \mu} \phi_{. \nu )} =
(1/2)\L(\Fo_\kappa{}^\mu \phi_{. \nu} + F_{\kappa \nu} \phio^{. \mu}\R)$.
Next, application of the covariant differentiation $\bdo$ defined in Sec\. V,
whose only nonvanishing connection coefficients are the
$\Gammao_\kappa{}^\mu{}_\lambda$ of Eqs\. (29), is signified by use of a
``$\, {}_: \,$'' {\it and} insertion of a ``$\; \degree \;$'' if none is
already present, provided that the field being differentiated is representable
in terms of the $e_\mu$, the $\omega^\mu$, and their tensor products alone,
with coefficients independent of $\zeta$ (a representability that passes on to
the differential field).  As examples,
$\bdo (d \phi) = \bdo (\phi_{. \kappa} \omega^\kappa) =
\phio_{. \kappa : \lambda} \omega^\lambda \otimes \omega^\kappa$,
$\bdo \Ao = \bdo \L(\Ao_\kappa \omega^\kappa\R) =
\Ao_{\kappa : \lambda} \omega^\lambda \otimes \omega^\kappa$,
$\;\;\;\; \bdo \L(\Go^{-1} \Ao\R) = \bdo \L(\Ao^\mu e_\mu\R) =
\Ao^\mu{}_{: \lambda} \omega^\lambda \otimes e_\mu$, and
$\bdo F = \bdo (F_{\kappa \lambda} \omega^\lambda \otimes \omega^\kappa) =
\Fo_{\kappa \lambda : \nu} \omega^\nu \otimes \omega^\lambda \otimes
\omega^\kappa$, where
$$
\align
\phio_{. \kappa : \lambda}
 &= \phi_{. \kappa \lambda} - \phi_{. \pi} \Gammao_\kappa{}^\pi{}_\lambda, \\
\Ao_{\kappa : \lambda}
 &= \Ao_{\kappa . \lambda} - \Ao_\pi \Gammao_\kappa{}^\pi{}_\lambda, \\
\Ao^\mu {}_{: \lambda}
 &= \Ao^\mu{}_{. \lambda} + \Ao^\pi \Gammao_\pi{}^\mu{}_\lambda,
\tag61 \\
\vspace{-10pt}
\intertext{and}
\vspace{-10pt}
\Fo_{\kappa \lambda : \nu}
 &= F_{\kappa \lambda . \nu}
    - F_{\kappa \pi} \Gammao_\lambda{}^\pi{}_\nu
    - F_{\pi \lambda} \Gammao_\kappa{}^\pi{}_\nu.
\endalign
$$
Because $\bdo \Go^{-1} = 0$, the raising of an index with the $\go^{\mu \nu}$
commutes with the ``$\; \degree \, {}_:$'' operation; for example,
$\Ao^\mu{}_{: \lambda} = \Ao_{\kappa : \lambda} \go^{\kappa \mu}$ because
$\bdo \L(\Go^{-1} \Ao\R) = \Go^{-1} \bdo \Ao$.  Finally, $\Thetao$,
$\Phio$, $\Psio$, and $\Eo$ stand for curvature fields built from $\bdo$ with
the help of $\Go$ and $\Go^{-1}$.  Specifically,
$\Thetao = \omega^\kappa \otimes \Thetao_\kappa{}^\mu e_\mu$,
$\Thetao_\kappa{}^\mu = 2 (d_\wedge \omegao_\kappa{}^\mu -
\omegao_\kappa{}^\pi \lwedge \omegao_\pi{}^\mu ) =
\Thetao_\kappa{}^\mu{}_{\lambda \nu} \omega^\nu \otimes \omega^\lambda$,
$\Psio = \omega^\kappa \otimes \Phio_{\kappa \lambda} \omega^\lambda$,
$\Psio = \Psio$, and
$\Eo = \omega^\kappa \otimes \Eo_{\kappa \lambda} \omega^\lambda$, where
$$
\align
\Thetao_\kappa{}^\mu{}_{\lambda \nu}
 &= 2 \L(\Gammao_\kappa{}^\mu{}_{[\lambda . \nu]}
             + \Gammao_\kappa{}^\pi{}_{[\lambda} \Gammao_\pi{}^\mu{}_{\nu]}
             + \Gammao_\kappa{}^\mu{}_\pi C_\lambda{}^\pi{}_\nu\R), \\
\Phio_{\kappa \lambda}
 &= \Thetao_\kappa{}^\rho{}_{\lambda \rho}, \\
\Psio
 &= \Phio_\pi{}^\pi,
\tag62 \\
\vspace{-10pt}
\intertext{and}
\vspace{-10pt}
\Eo_{\kappa \lambda}
 &= \Phio_{\kappa \lambda} - (1/2) \Psio \go_{\kappa \lambda}.
\endalign
$$
Because $\omegao_\kappa{}^d = \omegao_d{}^\mu = \omegao_d{}^d = 0$ and
$\Gammao_\kappa{}^\mu{}_{\lambda , d} = 0$, the only nonvanishing curvature
2-forms of $\bdo$ are the $\Thetao_\kappa{}^\mu$, so $\Thetao$ as given is the
curvature tensor field of $\bdo$.  In \text{space-time-\!-time} $\Thetao$ is
the curvature tensor field for the space-time metric $\Go$ picked out by the
gauge selection of the hypersurface $\eusm S$ on which to have $\zeta = 0$.

With these abbreviations all in place the detelescoped versions of $\Thetahat$
and $\Phihat$ are as shown in the Appendix.  From them one computes that
$$
\split
\Psihat = e^{-2 \zeta} \Psio
  &+ (d-2) e^{-2 \zeta} \L[2 \Ao^\rho{}_{: \rho}
                                      - (d-3) \Ao^\rho \Ao_\rho\R] \\
  &- 2 e^{-2 \zeta} \phi^{-1}
       \L[\phio^{. \rho}{}_{: \rho}
              - (d-3) \Ao^\rho\phi_{. \rho} \R] \\
  &+ \epshat (1/4) e^{-4 \zeta} \phi^2 \Fo_\pi{}^\rho \Fo_\rho{}^\pi
   - \epshat (d-1)d \, \phi^{-2},
\endsplit
\tag63
$$
and then that
$$
\align
\Ehat_{\kappa \lambda}
 &= \Eo_{\kappa \lambda}
{\topaligned
  &+ (d-3) \L[\Ao_{(\kappa : \lambda)}
                      - \Ao^\rho{}_{: \rho} \go_{\kappa \lambda}\R] \\
  &+ (d-3) \L[\Ao_\kappa \Ao_\lambda
                      + (1/2)(d-4) \Ao^\rho \Ao_\rho
                                            \go_{\kappa \lambda} \R] \\
  &- \phi^{-1} \L[\phio_{. \kappa : \lambda}
                      - \phio^{. \rho}{}_{: \rho} \go_{\kappa \lambda}\R] \\
  &- \phi^{-1}
          \L[2 \Ao_{(\kappa} \phi_{. \lambda )}
                + (d-4) \Ao^\rho \phi_{. \rho} \go_{\kappa \lambda} \R] \\
  &+ \epshat (1/2) e^{-2 \zeta} \phi^2
      \L[\Fo_\kappa{}^\rho F_{\rho\lambda}
             - (1/4) \Fo_\pi{}^\rho \Fo_\rho{}^\pi
                                    \go_{\kappa \lambda} \R] \\
  &+ \epshat (1/2)(d-2)(d-1) e^{2 \zeta} \phi^{-2}
                                                 \go_{\kappa \lambda},
\endtopaligned} \\
\vspace{-7pt}
\tag64 \\
\vspace{-7pt}
\Ehat_{\kappa d}
 &= \epshat (1/2) e^{-2 \zeta}
     \L[\phi \Fo_\kappa{}^\rho{}_{: \rho}
            - (d-5) \phi \Fo_\kappa{}^\rho \Ao_\rho
            + 3 \Fo_\kappa{}^\rho \phi_{. \rho} \R]
 + (d-2) \phi^{-2} \phi_{. \kappa}, \\ 
\Ehat_{d \lambda}
 &= \Ehat_{\lambda d}, \\
\vspace{-7pt}
\intertext{and}
\vspace{-7pt}
\Ehat_{dd}
 &= - \epshat (1/2) e^{-2 \zeta} \Psio
    - \epshat (d-2) e^{-2 \zeta}
       \L[\Ao^\rho{}_{: \rho} - (1/2)(d-3) \Ao^\rho \Ao_\rho \R] \\
 &\qquad - (3/8) e^{-4 \zeta} \phi^2 \Fo_\pi{}^\rho \Fo_\rho{}^\pi
         + (1/2)(d-2)(d-1) \phi^{-2}.
\endalign
$$

In the ancestral Kaluza geometry much of the complexity in these expressions
goes away, taking with it many of the possibilities for interactions among the
various fields.  (For the sake of comparison the corresponding expressions for
the Kaluza geometry are presented at the end of the Appendix.)
\vskip 15pt

\noindent {\bf X.  Residual Curvature}
\vskip 10pt

An important concept specific to the geometry of conformally constrained
metrics is that of residual curvature.  Loosely, the residual curvature is
what remains of the usual curvature when the instruments used to measure
it shrink to infinitesimal size --- the curvature seen by a vanishingly
small observer, so to speak.  A little less loosely, it is the limiting 
curvature at the ends of the trajectories of $\xi$ where the conformal
factor in $G = e^{2 \zeta} \Go$ becomes infinite.  The notion of residual
curvature does not apply to Kaluza metrics, which, being {\it isometrically}
constrained, have no factor $e^{2 \zeta}$ and therefore cannot have
$e^{2 \zeta} \to \infty$.  It requires for its definition that translations
along $\xi$ generate actual expansion of the metric $G$.  Moreover, $\eusm M$
needs to be $\xi$-complete, in order that along each $\xi$-path the
integration-parameter coordinate $\zeta$ might increase without bound
\cite{18}.

Consider on $\eusm M$ the frame system $\{e_\Mbar\}$ for which
$e_\mubar = e^{-\zeta} e_\mu$ and $e_\dbar = e_d$, with dual
$\{\omega^\Mbar\}$ given by $\omega^\mubar = e^\zeta \omega^\mu$ and
$\omega^\dbar = \omega^d$.  Referring to Eq\. ($4''$) one sees that in this
frame system
$$
\Ghat = \omega^\mubar \otimes \go_{\mubar \nubar} \omega^\nubar
         + \epshat \omega^\dbar \otimes \omega^\dbar,
\tag65
$$
where $\go_{\mubar \nubar} = \go_{\mu \nu}$.  From this it follows that
${\eusm L}_\xi \L(\Ghat e_\mubar e_\nubar\R) =
\partial \go_{\mubar \nubar} / \partial \zeta = 0$,
${\eusm L}_\xi \L(\Ghat e_\mubar e_\dbar\R) =
\partial 0 / \partial \zeta = 0$, and
${\eusm L}_\xi \L(\Ghat e_\dbar e_\dbar\R) =
\partial \epshat / \partial \zeta = 0$, in other words that all metrical
relationships determined by $\Ghat$ among the vector fields $e_\Mbar$ remain
fixed under translation along $\xi$.  The same of course holds true for the
covector fields $\omega^\Mbar$.  Beyond the normality of $e_d$ and the
orthogonality between the $e_\mu$ and $e_d$ the controlling fact here is that
the $e_\mu$ are Lie constant along $\xi$, which entails that
${\eusm L}_\xi e_\mubar = -e_\mubar$, hence that
${\eusm L}_\xi \L(\Ghat e_\mubar e_\nubar\R) =
\L({\eusm L}_\xi \Ghat\R) e_\mubar e_\nubar +
\Ghat \L({\eusm L}_\xi e_\mubar\R) e_\nubar +
\Ghat e_\mubar \L({\eusm L}_\xi e_\nubar\R) =
2 G e_\mubar e_\nubar - 2 \Ghat e_\mubar e_\nubar =
2 e^{-2 \zeta } \L(G e_\mu e_\nu - \Ghat e_\mu e_\nu\R) = 0$.

If $T$ is a tensor field of $\eusm M$, we can expand $T$ in terms of the
$e_\Mbar$ and the $\omega^\Mbar$, then can ask whether the components of
$T$ in this expansion have limits as $\zeta \to \infty$.  If all do, then the
tensor field $T_\infty$ whose components in $\{e_\Mbar\}$ are these limits
is to be called the ``residual'' of $T$.  More precisely, suppose that $T$ is
a tensor field of $\eusm M$, that $\tbar$ is a component of $T$ in
$\{e_\Mbar\}$ (and $\{\omega^\Mbar\}$), and that $P$ is a point in the 
domain of $T$.  Let $Q$ be a point lying on the trajectory of $\xi$ through
$P$ and free to move along it.  Let $\tbar_\infty (P) := \lim \tbar (Q)$ if
this limit exists as $Q$ moves along the trajectory so that
$\zeta (Q) \to \infty$; if the limit does not exist, then assign no meaning to
$\tbar_\infty (P)$.  If $\tbar_\infty (P)$ thus defined exists for each
such component $\tbar$ and point $P$, then the tensor field whose components in
$\{e_\Mbar\}$ are the corresponding scalar fields $\tbar_\infty$ is called
the {\bf residual of} $T$ and is denoted by $T_\infty$.  Briefly put, if, for
example,
$T = \omega^K \lotimes T_K{}^M\!{}_L \omega^L \lotimes e_M =
\omega^\Kbar \lotimes T_\Kbar{}^\Mbar\!{}_\Lbar \omega^\Lbar \lotimes
e_\Mbar$, then
$T_\infty := \omega^\Kbar \lotimes
\L(\lim_{\zeta \to \infty} T_\Kbar{}^\Mbar\!{}_\Lbar\R)
\omega^\Lbar \lotimes e_\Mbar$.

As an illustration,
$\Ghat = \omega^\mubar \otimes \ghat_{\mubar \nubar} \omega^\nubar +
\epshat \omega^\dbar \otimes \omega^\dbar$, where
$\ghat_{\mubar \nubar} = e^{-2 \zeta} g_{\mu \nu} = \go_{\mu \nu}$,
and therefore $\Ghat_\infty = \Ghat$, inasmuch as
$\lim_{\zeta \to \infty} \ghat_{\mubar \nubar} =
\lim_{\zeta \to \infty} \go_{\mu \nu} = \go_{\mu \nu} =
\ghat_{\mubar \nubar}$ and $\lim_{\zeta \to \infty} \epshat = \epshat$.
Similarly,
$\L(\Ghat^{-1}\R){}_\infty = \Ghat^{-1} =
\L(\Ghat_\infty\R)^{-1}$ and $A_\infty = A$.  On the other hand
$\Go = \omega^\mubar \otimes \go_{\mubar \nubar} \omega^\nubar$ with
$\go_{\mubar \nubar} = e^{-2 \zeta} \go_{\mu \nu}$, and
$\lim_{\zeta \to \infty} \L(e^{-2 \zeta} \go_{\mu \nu}\R) = 0$, so
$\Go_\infty = 0$.  By the same token $\Ao_\infty = 0$ and $F_\infty = 0$.
But $\Go^{-1} = e_\mubar \otimes \go^{\mubar \nubar} e_\nubar$, where
$\go^{\mubar \nubar} = e^{2 \zeta} \go^{\mu \nu}$, and if
$\go^{\mu \nu} \neq 0$, then $e^{2 \zeta} \go^{\mu \nu}$ has no limit as
$\zeta \to \infty$, so $\L(\Go^{-1} \R){}_\infty$ is not defined.

Several observations can be made:
1)~The component $\tbar$ of $T$ in $\{e_\Mbar\}$ is related to the
corresponding component $t$ of $T$ in
$\{e_M\}$ by $\tbar = e^{(\alpha - \beta) \zeta} t$, where $\alpha$ is the
number of contravariant, $\beta$ the number of covariant indices of $t$ that
differ from $d$; thus $\lim_{\zeta \to \infty} \tbar$ can equally well be
calculated as
$\lim_{\zeta \to \infty} \L(e^{(\alpha - \beta) \zeta} t\R)$.
2)~The residual of $T$, though defined in a particular gauge, is in fact gauge
invariant:  if $\zeta' = \zeta - \lambda$ with
$\partial \lambda / \partial \zeta = 0$, $e_{\mubar'} = e^{-\zeta'} e_\mu$,
$e_{\dbar^{'}} = e_d$, $\omega^{\mubar'} = e^{\zeta'} \omega^\mu$,
$\omega^{\dbar'} = \omega^d$, and, for example again,
$T = \omega^K \lotimes T_K{}^M\!{}_L \omega^L \lotimes e_M$, then
$T_{\kappabar'}{}^{\mubar'}\!{}_{\lambdabar^{'}} =
e^{-\zeta'} T_\kappa{}^\mu{}_\lambda =
e^\lambda e^{-\zeta} T_\kappa{}^\mu{}_\lambda =
e^\lambda T_\kappabar{}^\mubar{}_\lambdabar$, so
$\omega^{\kappabar'} \lotimes
\L(\lim_{\zeta' \to \infty}
T_{\kappabar'}{}^{\mubar'}\!{}_{\lambdabar^{'}}\R)
\omega^{\lambdabar'} \lotimes e_{\mubar'} =
\omega^\kappabar \otimes e^{-\lambda}
\L(\lim_{\zeta' \to \infty}
T_{\kappabar'}{}^{\mubar'}\!{}_{\lambdabar^{'}}\R)
\omega^\lambdabar \otimes e_\mubar =
\omega^\kappabar \otimes
\L(\lim_{\zeta \to \infty} T_\kappabar{}^\mubar{}_\lambdabar\R)
\omega^\lambdabar \otimes e_\mubar$; this together with analogous results
for the other components of $T$ expresses the gauge invariance of $T_\infty$.
3)~The definition of $T_\infty$ is also independent of the choice of the
adapted frame system $\{e_\mu , e_d\}$, a consequence of the fact that if
$\{e_{\mu''} , e_{d''}\}$ is another such system, then $e_{d''} = e_d$ and
$e_{\mu''} = J_{\mu''}{}^\mu e_\mu$ with $J_{\mu''}{}^\mu =
J_{\mu''}{}^{\mu'} J_{\mu'}{}^\mu$, which is independent of $\zeta$
(cf\. Sec\. V).
4)~The components of $T_\infty$ in the frame system $\{e_\Mbar\}$ are
constant on each trajectory of $\xi$.
5)~If the metric is conformally constrained in two directions (two $\xi$'s,
with ${\eusm L}_\xi \Ghat = 2 G$ for each), then each constraint produces a
residual of $T$, and these can differ; strictly, then one should qualify
{\it the} residual of $T$ by citing the generating vector field $\xi$.
6)~Algebraic symmetries of $T$ are preserved in $T_\infty$.
7)~Residuals of algebraic derivates of tensor fields (sums, products,
contractions, and the like) are the corresponding derivates of the residuals
of the constituents.

Now let us see what the residual of the curvature tensor field $\Thetahat$ is.
Noting that $\Thetahat_\kappa{}^\mu{}_{\lambda \nu}$ has one contravariant and
three covariant indices distinct from $d$, we conclude in light of observation
(1) above that $\Thetahat_\kappabar{}^\mubar{}_{\lambdabar \nubar} =
e^{-2 \zeta} \Thetahat_\kappa{}^\mu{}_{\lambda \nu}$.  Then, referring to the
first of Eqs\. (A.3) in the Appendix and multiplying both its members by
$e^{-2 \zeta}$, we see that as $\zeta \to \infty$ the only term on the right
that is not extinguished by an exponential factor is the last, and from this
it follows that
$$
\align
\L(\Thetahat_\infty\R){}_\kappabar{}^\mubar{}_{\lambdabar \nubar} 
 &= - \epshat 2 \phi^{-2} \go_{\kappa [\lambda} \go^\mu{}_{\nu]}.
\tag66 \\
\vspace{-7pt}
\intertext{Similar considerations show that}
\vspace{-7pt}
\L(\Thetahat_\infty\R){}_\kappabar{}^\dbar{}_{\lambdabar \dbar} 
 &= - \epshat \phi^{-2} \go_{\kappa \lambda}
  = - \L(\Thetahat_\infty\R){}_\kappabar{}^\dbar{}_{\dbar \lambdabar}
\tag67 \\ 
\vspace{-7pt}
\intertext{and}
\vspace{-7pt}
\L(\Thetahat_\infty\R){}_\dbar{}^\mubar{}_{\dbar \nubar} 
 &= - \phi^{-2} \go^\mu{}_\nu
  = - \L(\Thetahat_\infty\R){}_\dbar{}^\mubar{}_{\nubar \dbar},
\tag68
\endalign
$$
and that all remaining components of $\Thetahat_\infty$ vanish.  We
find, therefore, that
$$
\align
\Thetahat_\infty
 &= 2\phi^{-2} \L[\omega^\kappa \otimes
                      \L(\epshat g_{\kappa \lambda} \omega^\lambda \lwedge
                                          \omega^\mu\R) \otimes e_\mu \\
 &\qquad\qquad\qquad + \omega^\kappa \otimes
                      \L(\epshat g_{\kappa \lambda} \omega^\lambda \lwedge
                                            \omega^d\R) \otimes e_d
                     + \omega^d \otimes \L(\omega^d \lwedge \omega^\mu\R)
                                                           \otimes e_\mu\R],
\tag69 \\
\vspace{-7pt}
\intertext{which reduces to simply}
\vspace{-7pt}
\Thetahat_\infty
 &= \epshat 2 \phi^{-2} \L[\omega^K \lotimes
                              \L(\ghat_{KL} \omega^L \lwedge \omega^M \R)
                                                           \otimes e_M\R] \\ 
\vspace{-7pt}
\tag70 \\
\vspace{-7pt}
 &= \epshat \phi^{-2} \L[\omega^K \lotimes \ghat_{KL}
                            \L(\omega^L \lotimes \omega^M
                                   - \omega^M \lotimes \omega^L\R)
                                                           \otimes e_M\R].
\endalign
$$
This, then, is the residual of the curvature tensor field of $\Ghat$, or, a
little more succinctly, the {\bf residual curvature tensor field of} $\Ghat$.

One computes easily the {\bf residual} (of the) {\bf contracted curvature
tensor field of} $\Ghat$ by computing the corresponding contraction of
$\Thetahat_\infty$:
$$
\align
\Phihat_\infty
 &= - \epshat (d-1) \phi^{-2} \Ghat;
\tag71 \\
\vspace{-7pt}
\intertext{the {\bf residual} (of the) {\bf curvature scalar field of}
$\Ghat$ by computing the trace of $\Ghat^{-1} \Phihat_\infty$:}
\vspace{-7pt}
\Psihat_\infty
 &= - \epshat (d-1)d \; \phi^{-2};
\tag72 \\
\vspace{-7pt}
\intertext{and the {\bf residual} (of the) {\bf Einstein tensor field of}
$\Ghat$ by computing $\Phihat_\infty - (1/2) \Psihat_\infty \Ghat_\infty$:}
\vspace{-7pt}
\Ehat_\infty
 &= \epshat (1/2)(d-2)(d-1) \phi^{-2} \Ghat.
\tag73
\endalign
$$
It is also easy to learn that the residuals $\Thetao_\infty$ and $\Phio_\infty$
of the curvature fields $\Thetao$ and $\Phio$\ of $\bdo$ both vanish, that
$\L(\Go^{-1} \Phio\R){}_\infty = \Go^{-1} \Phio$ and
$\Psio_\infty = \Psio$, and that $\Eo_\infty$ vanishes.

Comparing Eq\. (69) and Eqs\. (A.3), we see that the components of
$\Thetahat_\infty$ in $\{e_\mu , e_d\}$ are just those additive contributions
to the components of $\Thetahat$ that do not depend on $\Ao_\mu$\ or on any
derivative of $\phi$, $\Ao_\mu$, or $\go_{\mu \nu}$.  In the case of the
prototypically conformally constrained de Sitter and hyper-de Sitter metrics
of Eqs\. (1) and (2) the latter quantities all vanish, leaving only the
constants $\phi$, and $\go_{\mu \nu}$ to determine the curvature fields.
Consequently, for these metrics $\Thetahat = \Thetahat_\infty$ as it is
expressed in Eqs\. (69) and (70), but particularized by the specialization of
the $\go_{\kappa \lambda}$ and by the fact that $\phi^{-2} = 1/R^2$.  The
manifolds with these metrics are, as previously remarked, open submanifolds of
hyperboloidal ``spheres'' of radius $R$; they have, therefore, uniform
sectional curvature of magnitude $1/R^2$, uniform at each point with respect to
choice of section (isotropic, in other words), and uniform from point to point.

In the general case there is no such uniformity of ordinary sectional
curvature.  Residual sectional curvature, however, is always isotropic, and is
uniform if $\phi$ is constant.  If $a$ and $b$ are a pair of tangent vectors at
the point $P$ of $\eusm M$, then from Eq\. (70) it follows readily that,
at $P$,
$$
\L(\Ghat \Thetahat_\infty\R) abab =
                - \epshat \phi^{-2} \l[\L(\Ghat aa\R)\L(\Ghat bb\R)
                - \L(\Ghat ab\R)^2\r].
\tag74
$$
This equation implies that if the square
$\L(\Ghat (P) aa\R)\L(\Ghat (P) bb\R) - \L(\Ghat (P) ab\R)^2$
of the area of the bivector $a \wedge b$ is not 0, then the {\bf residual
sectional curvature of $\Ghat$ at $P$ in the direction of $a \wedge b$},
defined in complete analogy with the ordinary sectional curvature as the
fraction of that square that the number $\L(\Ghat \Thetahat_\infty\R)(P)abab$
comes to, is $-\epshat \phi^{-2} (P)$.  As this is independent of $a$ and $b$,
the residual sectional curvature is isotropic at $P$; clearly it is uniform
from point to point only if $\phi$ is constant.  Even when not uniform,
however, it is constant on each trajectory of $\xi$, simply because
${\eusm L}_\xi \phi = 0$.  A shorter way of stating the facts is to say
that $\Ghat$ is {\bf residually spherical} at each point of $\eusm M$, with
vertically uniform {\bf residual radius of curvature} $\phi$ and
{\bf residual curvature} $-\epshat \phi^{-2}$.

We have seen that the $\Ghat$ lengths of the horizontal vectors $e_\mubar$ in
the reference frame $\{e_\Mbar\}$ used in calculating residual curvature stay
fixed as we push these vectors vertically along a trajectory of $\xi$.  The
other side of this is that their lengths as specified by the metric $\Go$ do
not stay fixed as we push them along.  In fact, $\Go e_\mubar e_\mubar =
e^{-2 \zeta} \Go e_\mu e_\mu = e^{-2 \zeta} \go_{\mu \mu}$, so
$\Go e_\mubar e_\mubar \to 0$ as $\zeta \to \infty$.  Also,
$\Go e_\dbar e_\dbar = \Go e_d e_d = 0$.  Thus the residual curvatures are
limits of ordinary curvatures measured against frames of vanishingly small
$\Go$ dimensions.

When $\Ghat$ is a \text{space-time-\!-time} metric, the $\Go$ lengths are the
usual dimensions of space and time, and it is in terms of these familiar
dimensions that the frame vectors $e_\mubar$ shrink to infinitesimal size as
$\zeta \to \infty$.  If we think of those frame vectors as abstract measuring
instruments belonging to a family of increasingly microscopic observers
stationed at a space-time event $\eusm E$, then the effect of their shrinking
is that the observers at the extreme microscopic end of the family are able to
perceive and to measure the curvatures of only their most immediate
surroundings, which to them are indistinguishable from a flat space-time region
embedded in a hyper-de Sitter, \text{space-time-\!-time} sphere of radius
$\phi (\eusm E)$.  Thus the residual curvatures, depending only on $\phi$,
represent an aspect of the geometry more infinitesimal in scale than that
represented by the nonresidual portions of curvatures embodied in the tensor
field $\Thetahat - \Thetahat_\infty$ and depending on $\Ao_\mu$ and the
derivatives of $\phi$, $\Ao_\mu$, and $\go_{\mu \nu}$ as well as on $\phi$ and
$\go_{\mu \nu}$ --- an ultralocal, as opposed to a merely local, aspect, one
could say.  This distinction between the local and the ultralocal aspects comes
into play when field equations for \text{space-time-\!-time} are to be derived
from an action principle.  By adopting for the action density the nonresidual
portion $\Psihat - \Psihat_\infty$ of the curvature scalar field, one can favor
\text{space-time-\!-times} that extremize not total curvature, but the total
deviation of curvature from the isotropic, ultralocal, vacuum emulating
residual curvature \cite{19}.
\vskip 15pt

\noindent {\bf XI.  Space-Time-\!-Time}
\vskip 10pt

The de Sitter metric $\Ghat$ of Eq\. (1) and the manifold $\eusm M$ on which it
is defined arise out of ordinary three-dimensional Euclidean space through the
following construction \cite{1}: the point of $\eusm M$ whose address is
$\[[ x,y,z,t \]]$ is (identified with) the Euclidean sphere of radius $r$
centered at $\[[ x,y,z \]]$, where $t := - \ln (r/R)$; if this sphere and an
infinitesimally neighboring sphere of radius $r + dr$ centered at
$\[[ x+dx , y+dy, z+dz \]]$ miss being tangent to one another by the angular
amount $d \alpha$ (the radian measure of their angle of intersection if the
spheres meet), then the squared distance between the corresponding points of
$\eusm M$ is (and this defines $\Ghat$) the number $R^2 d \alpha^2$.

The same construction applied to Minkowski space-time (which for present
purposes is interchangeable with de Sitter space-time, being conformally
equivalent to it and therefore having matching spheres and angles of 
intersection) yields both of the hyper-de Sitter metrics $\Ghat_-$ and
$\Ghat_+$ of Eq\. (2).  The metric $\Ghat_-$ results when the Minkowski spheres
(three-dimensional hyper-hyperboloids of revolution, in the Euclidean sense)
are of the one-sheeted variety, their points lying in spacelike directions from
their centers.  When the spheres are those of the two-sheeted variety, whose
points lie in timelike directions from their centers, $\Ghat_+$ results.  In
either case the sphere of radius $s$ centered at $\[[ x,y,z,t \]]$ corresponds
to the point (of ${\eusm M}_-$ or of ${\eusm M}_+$) with address
$\[[ x,y,z,t, \zeta \]]$, where $\zeta := - \ln (s/R)$.  And in either case the
squared distance between neighboring points is $R^2 d \beta^2$, where $d \beta$
is the angular amount by which the corresponding neighboring spheres fall short
of tangency \cite{20}.

It is because of this shared construction, which in the iterated application
defines the new coordinate $\zeta$ in terms of the radius $s$ precisely in the
manner that when first applied it defines the new, temporal coordinate $t$ in
terms of the radius $r$, {\it and on no other ground}, that I have attached the
label \text{space-time-\!-time} to manifolds with conformally constrained
metrics modeled on $\Ghat_-$ or on $\Ghat_+$.  That the signature $+++--$ of
$\Ghat_-$ appears to fit the label and the signature $+++-+$ of $\Ghat_+$
appears not to do so is of no consequence, for in either case the coordinate
$\zeta$ represents a geometrical entity thoroughly comparable to the entity
that the coordinate $t$ represents, justifiably called a ``time'' --- but a
time of a higher order, of course.  A fair description of the situation would
be that $t$ is ``space's time'' and $\zeta$ is ``space-time's time'' \cite{21}.

There being no geometrical reason to prefer the one kind of Minkowski sphere
to the other, it seems a half-measure to model physical systems by use of
conformally constrained metrics bearing either one of these signatures, to
the exclusion of those bearing the other, or to use one today and the other
tomorrow.  Expansion of the geometry to include the two signatures on equal
footing urges itself as an essential further step.  One way to effect such an
expansion is to complexify the secondary time coordinate $\zeta$, and along
with it the scalar field $\phi$ and the electromagnetic potential field $\Ao$.
When that is done, new elements become available for physical interpretation.
Conspicuous among them are 1)~a reciprocal coupling between pure imaginary
gauge transformations of $\Ao$ and complex phase shifts of $\phi$, and 2)~an
investing of each geodesic with a varying complex phase rotation whose
frequency parameter adjusts to the environs of the geodesic.  These particular
elements beg to be linked up with quantum mechanical phase phenomena, and that
rather clearly demands the forging of a link between the geometrical field
$\phi$ and the electron wave field (Schr\"odinger's $\psi$) of quantum theory.
The forging of such a link will, I believe, allow one ultimately to say, not
that geometry has been quantized, but that the quantum has been geometrized.
\vskip 15pt

\noindent {\bf APPENDIX.  Curvature Components}
\vskip 10pt

From the curvature 2-forms $\Thetahat_K{}^M$ as expressed in Eqs\. (51) one
readily extracts the curvature components $\Thetahat_K{}^M\!{}_{LN}$ that occur
in $\Thetahat_K{}^M = \Thetahat_K{}^M\!{}_{LN} \, \omega^N \lwedge \omega^L =
\Thetahat_K{}^M\!{}_{LN} \, \omega^N \lotimes \omega^L$.  Those that do not
vanish identically are given by
$$
\align
\Thetahat_\kappa{}^\mu{}_{\lambda \nu}
 &= \Theta_\kappa{}^\mu{}_{\lambda \nu}
    - \epshat \phi^2 F_\kappa{}^{(\mu} F_{[\lambda) \nu]}
    - \epshat 2 \phi^{-2} g_{\kappa [\lambda} g^\mu{}_{\nu]}, \\
\Thetahat_\kappa{}^\mu{}_{d \nu}
 &= \epshat (1/2)\! \l(\phi F_\kappa{}^\mu{}_{; \nu}
    + 2 F_{(\kappa}{}^\mu \phi_{. \nu)}
    + 2 F_\kappa{}^{(\mu} \phi_{. \nu)}\r)
    + 2 \phi^{-2} \phi_{. [\kappa} g^{\mu]}{}_\nu, \\
\Thetahat_\kappa{}^d{}_{\lambda \nu}
 &= \phi F_{\kappa [\lambda ; \nu]}
    - 2 \phi_{. (\kappa} F_{[\lambda) \nu]}
    + \epshat 2 \phi^{-2} g_{\kappa [\lambda} \phi_{. \nu]}, \\
\Thetahat_d{}^\mu{}_{\lambda \nu}
 &= - \epshat g^{\mu \kappa} \Thetahat_\kappa{}^d{}_{\lambda \nu},
\tag"(A.1)" \\
\Thetahat_\kappa{}^d{}_{\lambda d}
 &= - \phi^{-1} \phi_{, \kappa ; \lambda}
    - \epshat (1/4) \phi^2 F_\kappa{}^\pi F_{\pi \lambda}
    - \epshat \phi^{-2} g_{\kappa \lambda}, \\
\vspace{-7pt}
\intertext{and}
\vspace{-7pt}
\Thetahat_d{}^\mu{}_{d \nu}
 &= \epshat g^{\mu \kappa} \Thetahat_\kappa{}^d{}_{\nu d},
\endalign
$$
and the antisymmetry $\Thetahat_K{}^M\!{}_{NL} = -\Thetahat_K{}^M\!{}_{LN}$.
Here, in accordance with Eq\. (52),
$$
\align
\Theta_\kappa{}^\mu{}_{\lambda \nu} :\!@!@!@!@!@!@!@!
 &= 2 \l(\Gamma_\kappa{}^\mu{}_{[\lambda , \nu]}
     + \Gamma_\kappa{}^\pi{}_{[\lambda} \Gamma_\pi{}^\mu{}_{\nu]}
     + \Gamma_\kappa{}^\mu{}_\pi C_\lambda{}^\pi{}_\nu\r) \\
 &\qquad\quad - \! \l(g_\kappa{}^\mu F_{\lambda \nu}
                          - F_{\kappa [\lambda} g^\mu{}_{\nu]}
                          - g_{\kappa [\lambda} F^\mu{}_{\nu]}\r).
\tag"(A.2)"
\endalign
$$

When the righthand members of Eqs\. (A.1) are ``detelescoped'' by use of
Eqs\. (25)--(29), there results that
$$
\align
\Thetahat_\kappa{}^\mu{}_{\lambda \nu}
 &= \Thetao_\kappa{}^\mu{}_{\lambda \nu}
{\topaligned
  &+ 2 \! \l(\Ao_{(\kappa : [\lambda)} \go^\mu{}_{\nu]}
                 + \go_{\kappa [\lambda} \Ao^{(\mu}{}_{: \nu)]}\r) \\
  &+ 2 \! \l(\Ao_\kappa \Ao_{[\lambda} \go^\mu{}_{\nu]}
                 + \go_{\kappa [\lambda} \Ao^\mu \Ao_{\nu]}
                 - \Ao^\pi \Ao_\pi \go_{\kappa [\lambda} \go^\mu{}_{\nu]}\r) \\
  &- \epshat e^{-2 \zeta} \phi^2 \Fo_\kappa{}^{(\mu} F_{[\lambda) \nu]}
   - \epshat 2 e^{2 \zeta}
                      \phi^{-2} \go_{\kappa [\lambda} \go^\mu{}_{\nu]},
\endtopaligned} \\
\Thetahat_\kappa{}^\mu{}_{d \nu}
 &= \epshat (1/2) e^{-2 \zeta}
                     \l[\phi \Fo_\kappa{}^\mu{}_{: \nu}
                         + 2 \Fo_{(\kappa}{}^\mu \phi_{.\nu)}
                         + 2 \Fo_\kappa{}^{(\mu} \phi_{. \nu)}\r. \\
 &\qquad\qquad\qquad\qquad \l. 
    + 2 \phi \! \l(\Fo_{(\kappa}{}^\mu \Ao_{\nu)}
    + \Fo_\kappa{}^{(\mu} \Ao_{\nu)}
    - \Fo_{[\kappa}{}^\pi \Ao_\pi \go^{\mu]}{}_\nu\r)\r] \\
 &\qquad + 2 \phi^{-2} \phi_{. [\kappa} \go^{\mu]}{}_\nu, \\
\Thetahat_\kappa{}^d{}_{\lambda \nu}
 &= \phi \Fo_{\kappa [\lambda : \nu]} - 2 \phi_{. (\kappa} F_{[\lambda) \nu]}
      - 2 \phi \! \l(\Ao_{(\kappa} F_{[\lambda) \nu]}
                         - \go_{\kappa [\lambda} \Ao^\pi F_{\pi \nu]}\r)
\tag"(A.3)" \\
 &\qquad + \epshat 2 e^{2 \zeta} \phi^{-2}
                                      \go_{\kappa [\lambda} \phi_{. \nu]}, \\
\Thetahat_d{}^\mu{}_{\lambda \nu}
 &= - \epshat e^{-2 \zeta}
                 \go^{\mu \kappa} \Thetahat_\kappa{}^d{}_{\lambda \nu}, \\
\Thetahat_\kappa{}^d{}_{\lambda d}
 &= - \phi^{-1} \! \l[\phio_{. \kappa : \lambda}
               + \! \l(2 \Ao_{(\kappa} \phi_{. \lambda)}
                           - \go_{\kappa \lambda} \Ao^\pi \phi_{. \pi}\r)\r] \\
 &\qquad- \epshat (1/4) e^{-2 \zeta} \phi^2 \Fo_\kappa{}^\pi F_{\pi\lambda}
        - \epshat e^{2 \zeta} \phi^{-2} \go_{\kappa \lambda}, \\
\vspace{-7pt}
\intertext{and}
\vspace{-7pt}
\Thetahat_d{}^\mu{}_{d \nu}
 &= \epshat e^{-2 \zeta} \go^{\mu \kappa} \Thetahat_\kappa{}^d{}_{\nu d}.
\endalign
$$
The abbreviations that appear in these expressions are explained in Sec\. IX.

From Eqs\. (A.3) and the relation $\Phihat_{KL} = \Thetahat_K{}^R\!{}_{LR}$
it now follows that
$$
\align
\Phihat_{\kappa \lambda}
 &= \Phio_{\kappa \lambda}
{\topaligned
  &+ \! \l[(d-3) \Ao_{(\kappa : \lambda)}
               + \Ao^\rho{}_{: \rho} \go_{\kappa \lambda}
               + (d-3) \! \l(\Ao_\kappa \Ao_\lambda
                             + \Ao^\rho \Ao_\rho \go_{\kappa \lambda}\r)\r] \\ 
  &- \phi^{-1} \phio_{. \kappa : \lambda}
   - 2 \phi^{-1} \!
            \l[\Ao_{(\kappa} \phi_{. \lambda)}
               - (1/2) \Ao^\rho \phi_{. \rho} \go_{\kappa \lambda}\r] \\
  &+ \epshat (1/2) e^{-2 \zeta} \phi^2 \Fo_\kappa{}^\rho F_{\rho \lambda}
   - \epshat (d-1) e^{2 \zeta} \phi^{-2} \go_{\kappa \lambda},
\endtopaligned} \\
\Phihat_{\kappa d}
 &= \epshat (1/2) e^{-2 \zeta} \!
      \l[\phi \Fo_\kappa{}^\rho{}_{: \rho}
         - (d-5) \phi \Fo_\kappa{}^\rho \Ao_\rho
         + 3 \Fo_\kappa{}^\rho \phi_{. \rho}\r]
 + (d-2) \phi^{-2} \phi_{. \kappa},
\tag"(A.4)" \\
\Phihat_{d \lambda}
 &= \Phihat_{\lambda d}, \\
\vspace{-7pt}
\intertext{and}
\vspace{-7pt}
\Phihat_{dd}
 &= - \epshat e^{-2 \zeta} \phi^{-1} \!
       \l[\phio^{. \rho}{}_{: \rho} - (d-3) \Ao^\rho \phi_{. \rho})\r]
 - (1/4) e^{-4 \zeta} \phi^2 \Fo_\pi{}^\rho \Fo_\rho{}^\pi
         - (d-1) \phi^{-2},
\endalign
$$
where $\Phio_{\kappa \lambda} = \Thetao{}_\kappa{}^\rho{}_{\lambda \rho}$,
as said in Sec\. IX.

$$
\frac{\phantom{MMMMMMMMMMMMMMM}}{\phantom{MMMMMMMMMMMMMMM}}
$$

Here, for the sake of comparison with Eqs\. (63) and (64), are the forms that
$\Psihat$, $\Ehat_{\kappa \lambda}$, $\Ehat_{\kappa d}$, $\Ehat_{d \lambda}$,
and $\Ehat_{d d}$ would take for the Kaluza geometry, whose metric is
obtained from Eqs\.~($4'$) by the replacement $e^{2 \zeta} \to 1$:
$$
\Psihat = \Psio - 2 \phi^{-1} \phio^{. \rho}{}_{: \rho}
                + \epshat (1/4) \phi^2 \Fo_\pi{}^\rho \Fo_\rho{}^\pi,
\tag"(A.5)"
$$
$$
\align
\Ehat_{\kappa \lambda}
 &= \Eo_{\kappa \lambda}
   - \phi^{-1} \L[\phio_{. \kappa : \lambda}
                      - \phio^{. \rho}{}_{: \rho} \go_{\kappa \lambda}\R]
   + \epshat (1/2) \phi^2
      \L[\Fo_\kappa{}^\rho F_{\rho\lambda}
             - (1/4) \Fo_\pi{}^\rho \Fo_\rho{}^\pi
                                    \go_{\kappa \lambda} \R], \\
\Ehat_{\kappa d}
 &= \epshat (1/2)
     \L[\phi \Fo_\kappa{}^\rho{}_{: \rho}
            + 3 \Fo_\kappa{}^\rho \phi_{. \rho} \R], \\
\Ehat_{d \lambda}
 &= \Ehat_{\lambda d},
\tag"(A.6)" \\
\vspace{-7pt}
\intertext{and}
\vspace{-7pt}
\Ehat_{dd}
 &= - \epshat (1/2) \Psio - (3/8) \phi^2 \Fo_\pi{}^\rho \Fo_\rho{}^\pi.
\endalign
$$
The further replacement $\phi \to 1$ yields the forms of $\Psihat$,
$\Ehat_{\kappa \lambda}$, $\Ehat_{\kappa d}$, $\Ehat_{d \lambda}$, and
$\Ehat_{d d}$ for the Kaluza--Klein geometry.
\vfill\eject


\Refs\nofrills{\noindent {\bf References}}

\ref \no 1
\by H. G. Ellis
\jour {\it Found. Phys.} \vol 4 \yr 1974 \pages 311--319
\moreref \paper \jour \paperinfo Erratum: \vol 5 \yr 1975 \page p\. 193;
partially reprised in Sec\. XI of the present paper.  N.B.  In
several places the editor of the journal substituted ``declaration'' for the
word ``manifesto'' used in the manuscript.  I would have preferred ``credo''
as a compromise
\endref

\ref \no 2
\by H. G. Ellis
\book Abstracts of Contributed Papers \publ 8th International
Conference on General Relativity and Gravitation, Univ\. of Waterloo, Waterloo,
Ont., Canada \yr 1977 \page 138.  %
In this abstract both a ``weak form'' and a ``strong
form'' conformality constraint are defined.  The present discussion refers to
a yet stronger form (to be defined in Sec\. II), which will be identified
simply as {\it the} conformality constraint, the earlier versions being no
longer in use
\endref

\ref \no 3
\by Th. Kaluza
\jour {\it S.--B. Preuss. Akad. Wiss.} \vol 1921 \pages 966--972.  %
The reference here is to the geometry with the ``cylinder condition'' as Kaluza
intended it, without the additional restriction that the Killing vector field
have constant length, appended later by Klein \cite{a} and by Einstein
\cite{b}.  This restriction conceivably lies implicit in Kaluza's stated and
unstated assumptions; if so, he did not recognize it
\endref
\hskip -12pt \hbox{
\ref \key a
\by O. Klein
\jour {\it Z. Physik} \vol 37 \pages 895--906 \yr 1926 \moreref \vol 46
\pages 188--208 \yr 1927
\endref}

\hskip -12pt \hbox{
\ref \key b
\by A. Einstein \jour {\it S.--B. Preuss. Akad. Wiss., Phys.--math. Kl.}
\vol 1927
\pages 23--30
\endref}

\parindent 15pt
\ref
\by \paper \jour \paperinfo
\hskip 15pt
The Kaluza paper and the first Klein paper appear in English translation
in {\it Modern Kaluza--Klein Theories} (vol\. 65 in the series Frontiers
in Physics), T. Appelquist, A. Chodos, and P. G. O. Freund, eds.
(Addison--Wesley, Menlo Park, California, 1987); a facsimile of the Kaluza
paper also is included
\endref

\ref \no 4
\by H. Weyl \jour {\it S.--B. Preuss. Akad. Wiss.} \vol 1918 \pages 465--480
\moreref \jour {\it Ann. d. Physik} \vol 59 \pages 101--133 \yr 1919
\endref

\ref \no 5
\by E. Schr\"odinger \book {\it Expanding Universes}
\publ Cambridge Univ. Press \publaddr Cambridge, U.K. \yr 1956 \pages 28--33
\endref

\ref \no 6
\by \paper \jour \paperinfo
Strict adherence to the meaning of ``conformal'' would classify metrics
$\Ghat$ for which ${\eusm L}_\xi \Ghat = 0$ as ``isometrically
conformally constrained'' if ${\eusm L}_\xi \Ghat = 0$ and those for which
${\eusm L}_\xi \Ghat = 2 G$ as ``nonisometrically conformally constrained'';
the less accurate ``isometrically constrained'' and ``conformally constrained'' have the advantage of brevity
\endref

\ref \no 7
\by Th. Kaluza \jour {\it loc. cit.}  %
Kaluza was not adamant about this choice, indeed seemed willing to let it go
the other way if by so doing he could overcome a ``very large'' difficulty
pointed out to him by Einstein (p. 971, last paragraph)
\endref

\ref \no 8
\by \paper \jour \paperinfo
My reasons for calling $\eusm M$ a \text{space-time-{\it time}} manifold can
be found in Ref\. 1; they are discussed here in Sec\. XI
\endref

\ref \no 9
\by \paper \jour \paperinfo
This supposition, introduced here to simplify the construction, is not
essentially restrictive, as cases in which it is false can be brought
under it by judicious use of covering manifolds
\endref

\ref \no 10
\by \paper \jour \paperinfo
Weyl himself was mainly responsible for this amnesia, through his discovery
of the tie between electromagnetic gauge transformations and electron wave
field phase shifts (cf\. {\it Space-Time-Matter} (Dover, New York, 1950),
Preface to the First American Printing, p\. v, and references cited there)
\endref

\ref \no 11
\by \paper \jour \paperinfo
All skew-symmetrizations signified by [  ] and symmetrizations signified by
(  ) are to be carried out on {\it two} indices only, namely the leftmost
and the rightmost indices enclosed
\endref

\ref \no 12
\by \paper \jour \paperinfo
These commutators include the factor 1/2, viz., $[u,v] := (1/2)(uv - vu)$
\endref

\ref \no 13
\by W. Ehrenberg and R. Siday \jour {\it Proc. Phys. Soc.} (London)
\vol B62 \pages 8--21 \yr 1949
\endref

\ref \no 14
\by Y. Aharonov and D. Bohm \jour {\it Phys. Rev.} \vol 115
\pages 485--491 \yr 1959
\endref

\ref \no 15
\by \paper \jour \paperinfo
The derivation is not quite straightforward, because the coframe system
$\{ \omega^\mu , \omega^d \}$ is not required to be holonomic.
The Euler equations turn out to be, after division by 2,
$(\ghat_{KL} \pdot^L)\dot{}\, - (1/2) \pdot^M \ghat_{MN,K} \pdot^N
+ C_K {}^M {}_L \pdot^L \ghat_{MN} \pdot^N = 0$, where
$K, L, M, N = 1, \ldots@! ,d$ and $C_K {}^M {}_L =
(d_\wedge \omega^M ) e_L e_K = \omega^M [e_K , e_L]$ 
\endref

\ref \no 16
\by \paper \jour \paperinfo
See for example Eq\. (2.41), p\. 62, in {\it Electrodynamics and Classical
Theory of Fields and Particles}, A. O. Barut (Macmillan, New York, 1964)
\endref

\ref \no 17
\by D. M. Chase \jour {\it Phys. Rev.} \vol 95 \yr 1954 \pages 243--246, %
obtained results that seem to imply some connection in Einstein--Maxwell
theory between the ratio of active charge to active mass and the ratio
of passive charge to passive mass
\endref

\ref \no 18
\by \paper \jour \paperinfo
A weaker condition would suffice, merely that $\eusm M$ be
$\xi$-{\it forwardly complete}, in other words that on each $\xi$-path
the integration parameter run to $\infty$ 
\endref

\ref \no 19
\by \paper \jour \paperinfo
The restricted field equations of Ref\. 2 are derived from such an action
principle.  N.B.  The equation there labeled $\delta \psi$) is wrong,
and should be replaced by $\Ao^\mu {}_{: \mu} - 3 \Ao^\mu A_\mu =
(1/2) \Ro - (3/4) K \psi^2 \Fo^\mu {}_\nu \Fo^\nu {}_\mu$.  These equations,
derived in a gauge chosen to simplify description of the integration region
$D^5$, are not fully gauge invariant, nor are they intended to be
\endref

\ref \no 20
\by \paper \jour \paperinfo
See Ref\. 1; Eq\. (4) there needs a minus sign before the $d \beta^2$.
See also R. L. Ingraham, {\it Nuovo Cimento} {\bf 46} B (1978), 16--32
\endref

\ref \no 21
\by \paper \jour \paperinfo
A point to be emphasized is that, irrespective of any label applied to it,
the dimension coordinatized by $\zeta$ differs so radically in character
from the three spatial dimensions that no attempt to make it ``unobservable''
is demanded.  Making it periodic by rolling up the trajectories of $\xi$ into
circles (after the practice of Kaluza--Klein theorists) would be in fact
incompatible with the conformality constraint and therefore impossible.  Cases
in which the $\xi$-trajectories return repeatedly to the same space-time cross
section, each return occurring at a new event, might usefully be contemplated,
however
\endref

\endRefs
\enddocument